\documentclass[preprint, superscriptaddress,pra,aps,showpacs]{revtex4-1}

\usepackage{graphicx}
\usepackage{subfigure}
\usepackage{multirow}
\usepackage{siunitx}
\usepackage[cmex10]{amsmath}
\newcommand{\argmin}{\operatornamewithlimits{arg\ min}}
\def\mathbi#1{\textbf{\em #1}}

\graphicspath{{fig/}}
 \DeclareGraphicsExtensions{.eps, .pdf, .png}

\begin{document}

\title{Monte Carlo validation of optimal material discrimination using spectral x-ray imaging}

\author{S. J. Nik}
\affiliation{Department of Physics and Astronomy, University of Canterbury, Christchurch, New Zealand.}

\author{R. S. Thing}
\affiliation{Institute of Clinical Research, University of Southern Denmark, Odense, Denmark.}

\author{R. Watts} 
\affiliation{UVM MRI Center for Biomedical Imaging, University of Vermont College of Medicine, Burlington, VT, USA.}

\author{T. Dale} \affiliation{BlueFern, University of Canterbury, Christchurch, New Zealand.}

\author{B. Currie} 
\affiliation{Medical Physics and Bioengineering Department, Christchurch Hospital, Christchurch, New Zealand.}

\author{J. Meyer}
\affiliation{Department of Radiation Oncology, University of Washington Medical Center, Seattle, WA, USA.}

\begin{abstract} 
The aim of this work was to develop a framework to validate an algorithm for determination of optimal material discrimination in spectral x-ray imaging.
%The validation of a previous work on the optimization of material discrimination in spectral x-ray imaging is reported. 
Using Monte Carlo (MC) simulations based on the BEAMnrc package, material decomposition was performed on the projection images of phantoms containing up to three materials. The simulated projection data was first decomposed into material basis images by minimizing the z-score between expected and simulated counts. Statistical analysis was performed for the pixels within the region-of-interest consisting of contrast material(s) in the MC simulations. 
With the consideration of scattered radiation and a realistic scanning geometry, the theoretical optima of energy bin borders provided by the algorithm were shown to have an accuracy of $\pm$2\,keV for the decomposition of 2 and 3 materials. Finally, the signal-to-noise ratio predicted by the theoretical model was also validated. The counts per pixel needed for achieving a specific imaging aim can therefore be estimated using the validated model.

\end{abstract}

%\keywords{Simulation methods and programs; Analysis and statistical methods}

\maketitle

\section{Introduction}
With the ability of acquiring multiple energy resolved images in a single acquisition, spectral CT imaging can be considered an expansion of dual energy CT. Photon counting detectors (PCDs) with energy discriminating abilities, such as the Medipix and XPAD detectors, have been built to achieve this \cite{PangaudXPAD3,BallabrigaMPX32011}. Energy discriminating PCDs are equipped with tunable pulse height discriminators within the electronics of the PCDs. Modern photon counting detectors are often equipped with several independent discriminators, with as many as 8 provided in the Medipix3 detectors \cite{BallabrigaMPX32011,WalshMPX3.1}. Data associated with a higher energy level can be subtracted from that of a lower energy to form data for an energy bin \cite{Schlomkaspectraldemo}. 

Based upon Alvarez and Macovski's \cite{AlvarezEnergySelective} technique of dual-energy imaging, the advent of spectral x-ray imaging has enabled three-component decomposition. Given the projection data, material decomposition can be realized by estimating the thicknesses or the areal densities of specific materials, prior to reconstruction. The benefits of spectral x-ray imaging in material identification have been ubiquitously demonstrated for medical \cite{Schlomkaspectraldemo,RoesslPreclinical,LeSegmentation} and security applications \cite{BeldjoudiIdetification}. Higher numbers of energy bins have been demonstrated to be beneficial in material quantification \cite{FreyMaterialDecomposition}. For a limited number of bins, the optimal arrangement of energy windows that maximizes the spectral information for material separation remains unclear. 

Material decomposition in this work is performed by minimizing the z-score between the measurements and the expected counts given by the Beer-Lambert equation. Based on this approach, a theoretical model of optimizing the spectral information has previously been developed by minimizing the uncertainties of thickness estimates \cite{NikOptimal}. The focus of this paper is to validate the minimization of confidence regions on material quantities under the influence of Poisson counting noise, scattered radiation and a realistic scanning geometry. The theoretical algorithm was also extended to predict the variances of material thicknesses, which enables the estimation of counts per pixel needed for an optimal material discrimination. 
A framework of Monte Carlo (MC) simulations for spectral imaging is presented and the previously established material decomposition method was applied on the simulated data to validate the extended theoretical model. 

\section{Background}\label{background}

The complete formulation of our optimization model for material discrimination by minimizing the z-score has been presented in \cite{NikOptimal} and will be summarized here briefly.

Consider the linear attenuation coefficients $\mu_{i}$ of material {\it i} as a result of Compton (incoherent) scattering and the photoelectric effect. 
%\begin{equation}
%	\centering
%		\mu_{i}(E) = \mu_{i_{Compton}}(E) + \mu_{i_{Photoelectric}}(E).
%		\label{eq:totalattenuation}
%\end{equation}
The number of photons, {\it N} between energies $E_l$ and $E_h$ after being transmitted through $i = 1,\dots, m$ materials, as governed by the Beer-Lambert equation is:
\begin{equation}
	%N(E_l, E_h,\mathbi{t}) = \int_{E_l}^{E_h}  N_0(E)e^{-{\sum_{i=1}^m \mu_{i}(E)t_{i} } }
	N(E_l, E_h,\mathbi{t}) = \int_{E_l}^{E_h} \! N_0(E)e^{-{\sum_{i=1}^m \mu_{i}(E)t_{i}}} \, dE,
	\label{eq:Nintegral}
\end{equation}		
where $N_0$ is the number of incident photons. $\mathbi{t}$ represents a set of thicknesses $t_i$ for $i = 1, \dots, m$ materials. 

Given the linear dependency of the material attenuation functions, only two materials can be decomposed, if the imaging object does not present any k-edges within the energies considered \cite{RoesslKedge,WangOptimalTh}. However, a third material with a k-edge within the detected x-ray spectrum can be discriminated with 3 or more spectroscopic measurements. %(see e.g. \cite{RoesslKedge,WangOptimalTh}). 
In the regime of spectral x-ray imaging, at least as many bins, {\it n}, as materials have to be fitted for the discrimination of {\it m} materials ($n \geq m$). Henceforth, it is assumed that photons are binned into a minimum of $n = 2$ energy bins, for the separation of at least $m=2$ materials. Photons are allocated into energy bin {\it k} for $k = 1, \dots, n$, where $E_{(l,k)}$ and $E_{(h,k)}$ are the low and high limits for bin {\it k}, respectively. The photon count in bin {\it k} is denoted $N_k$, where ${N_k}$ follows a Poisson distribution with a mean of $\lambda_k$; the standard deviation is $\sigma_k = \sqrt{\lambda_k}$. 

\begin{figure}[t]
  \centering
   \includegraphics[width=0.28\textwidth,  angle=-90]{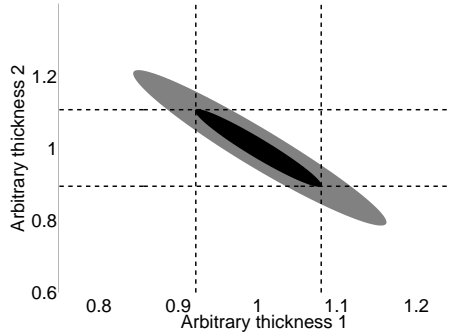}
   \caption{The black ellipse marks the 63\% confidence region formed by a z-score of unity. The outer ellipse represents the confidence region for a z-score of 2 for $m=2$ materials, encompassing a probability content of 98\%. Expanding this to $m=3$ materials results in a confidence volume. }
      \label{fig_confregBound}
\end{figure}

As $\lambda_k$ is sufficiently large, ${N_k}$ can be approximated to a Gaussian distribution. The z-score between the measurements, $\mathbi{x} = \{ x_k \}$, and the expected counts, $\boldsymbol{\lambda} = \{ \lambda_k \}$, can therefore be written as 
\begin{equation}
	z = \frac{\mathbi{x} - \boldsymbol{\lambda}} {\boldsymbol{\sigma}} = \frac{x_1 - \lambda_1}{\sqrt{\lambda_1}} = \frac{x - \lambda}{\sqrt{\lambda}},
	\label{eq:zscore}
\end{equation}
for measurements consisting of $n=1$ bin.
The Mahalanobis distance, which is the z-score for $n > 1$ energy bins, is given by \cite{RencherMVanalysis,NikOptimal}
\begin{equation}
%z &{}={}& \left\{ \sum_{k=1}^n  \left[ \frac{(x_k - \lambda_k)}{\sigma_k} \right]^2  \times \frac{1} {n} \right\}^\frac{1}{2} \nonumber \\
%&& =  { \left\{  \left[  \sum_{k=1}^n  \Big( x_k - \lambda_k(\mathbi{t}) \Big) ^2 \times \frac{1}{ \lambda_k( { \mathbi{t} } ) } \right]  \times \frac{1} {n} \right\}^{\frac{1}{2}}, }
z =  { \left\{  \left[  \sum_{k=1}^n  \Big( x_k - \lambda_k(\mathbi{t}) \Big) ^2 \times \frac{1}{ \lambda_k( { \mathbi{t} } ) } \right]  \times \frac{1} {n} \right\}^{\frac{1}{2}}, }
 \label{eq:zscorebins} 
\end{equation}
in which a factor of $1/n$ has been introduced for convenience to negate the dependency of {\it z} on the number of energy bins. Mapping the z-score in the thickness space therefore leads to an elliptical contour plot for $m=2$ materials and $n=2$ bins, indicating a multivariate normal distribution \cite{RencherMVanalysis,JamesStatsMethods}. A confidence region formed by a z-score of unity is shown as the black ellipse in figure \ref{fig_confregBound}, which contains a probability content, $\beta$, of 63\%. The $\beta$-value may be interpreted as meaning that there is a 63\% chance that given a measurement $\mathbi{x}$, the actual thicknesses would lie within this particular region. Similarly, the 98\% confidence region formed by a z-score of 2 is represented by the grey ellipse. Located at the center of the two-dimensional ellipse is a z-score of zero, corresponding to $\boldsymbol{\tau} = \{ \tau_i \}, i = 1,\dots, m$, where $\boldsymbol{\tau}$ is the combination of thicknesses that is most consistent with the measurement $\mathbi{x}$. The confidence ellipse can be expanded into any higher dimensions e.g. a volume for $m=3$ materials \cite{JamesStatsMethods,NikOptimal}. 

The bounding box of the ellipsoidal confidence region, as depicted in figure \ref{fig_confregBound}, enables the calculation of the standard deviations ($\sigma_i$) and correlation coefficient ($\rho$) of the thicknesses for the formation of the covariance matrix of the thickness population, $V$ \cite{JamesStatsMethods}:  
\[
V =
\left( {\begin{array}{ccc}
 \sigma_1^2 & \rho \sigma_1 \sigma_2 & \rho \sigma_1 \sigma_ 3 \\
 \rho \sigma_1 \sigma_2  & \sigma_2^2 &    \rho \sigma_2 \sigma_3          \\
 \rho \sigma_1 \sigma_3 &   \rho \sigma_2 \sigma_3   & \sigma_3^2 \\
 \end{array} } \right).
\] 
The diagonal elements in the matrix can be used to quantify the confidence region and thus the uncertainties of the thickness estimates. 
Given the number of energy bins {\it n}, the objective of the model is to locate the energy thresholds $E_{(l,k)}$ and $E_{(h,k)}$ for $k = 1,\dots, n$ that give the smallest confidence region in the thickness space, which was achieved by an exhaustive search through the space of all possible combinations of energy bins $E_{(l,k)}$ and $E_{(h,k)}$ in this paper and in the previous work \cite{NikOptimal}. 

\section{Methods}\label{methods}
Despite their promising potential, the performance of PCDs is at present limited by charge sharing \cite{BallabrigaMPX32011}, scattered radiation \cite{RoesslSensitivity}, finite energy resolution \cite{Schlomkaspectraldemo} and relatively low read-out speed \cite{RoesslPreclinical}. To investigate the achievable potential of spectral x-ray imaging, for example, Roessl {\it et al.} \cite{RoesslSensitivity} resorted to the ideal environment of CT simulations to investigate the maximum signal to noise ratio (SNR) in the basis images of high atomic number material to bypass the limitations. Other simulations of spectral x-ray imaging have been performed using commercial packages \cite{RoesslKedge,WeigelBreastCT,LengNoiseReduction}, open source packages \cite{GierschROSI,FreyMaterialDecomposition}, or analytical methods \cite{DucoteSim}. We chose a different MC simulation code system, known as BEAMnrc \cite{RogersBEAM,BEAMnrcManual}, because of its availability, ease of use as well as our previous experience with the system \cite{BrynThesis,RuneVirtualsCT}. The BEAMnrc system is based on the EGSnrc code \cite{EGSnrcManual} and comes with extensive documentation plus interactive graphical user interfaces. The recognition of the package through publication statistics and a review on the advantages on BEAMnrc over other MC packages was provided by Rogers\cite{RogersReview}. %including PENELOPE \cite{BaroPENELOPE}, MCNP \cite{BrownMCNP} and GEANT4 \cite{AgostinelliGEANT4} 

\subsection{Monte Carlo simulation setup}\label{ch4_met:MCsim}

\subsubsection{BEAMnrc simulations}\label{met:BEAMnrc}

\begin{figure}[t]
  \centering
   \includegraphics[width=0.55\textwidth]{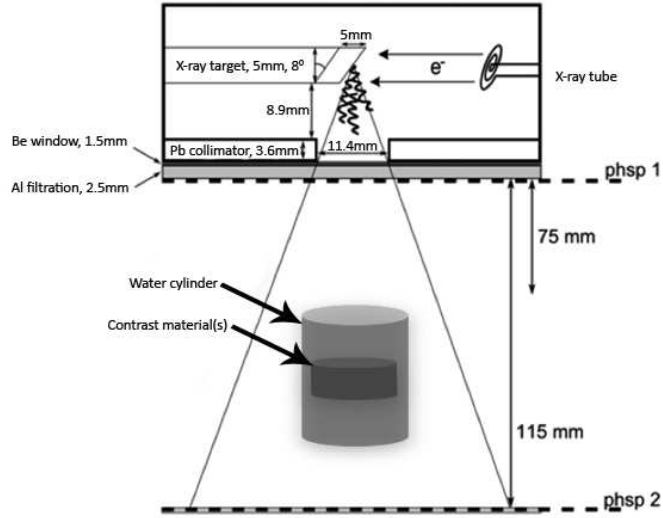}
    \caption{Simulation setup on BEAMnrc resembling the geometry of the locally developed Medipix All Resolution System (MARS) CT scanner (MARS Bioimaging Ltd, New Zealand) \cite{WalshMARSCT}. The phantom was designed to be contrast layer(s) embedded within a water cylinder for material decomposition of up to 3 materials. The distances from the exit window to the top of the imaging object and to the detector were defined to be 75\,mm and 115\,mm, respectively.} 
   \label{fig_BEAMsetup}
\end{figure} 

Using the BEAMnrc MC code system, simulations were carried out on the \href{http://www.bluefern.canterbury.ac.nz}{BlueFern\textsuperscript{\textregistered}} supercomputer at the University of Canterbury, Christchurch, New Zealand. %\footnote[1]{http://www.bluefern.canterbury.ac.nz}. 
The scanning geometry was set up to correspond to the locally built Medipix All Resolution System (MARS) CT scanner (MARS Bioimaging Ltd, New Zealand) \cite{WalshMARSCT}, as depicted in figure \ref{fig_BEAMsetup}. For the x-ray tube, CIRCAPP component module was used to replicate the round exit window and SLABS to include the 1.5mm beryllium and 2.5mm aluminum filtration corresponding to \cite{RoesslKedge}. The 90/10 atomic percent tungsten/rhenium alloy anode target was simulated with the XTUBE component module. The electron beam impingng on the target was simulated as a 120keV monoenergetic, parallel rectangular source energy incident from the side to enable validations of optimal energy bins with the previous work.

% In BEAMnrc, choices of component modules consist of cylinders, slabs and polygons \cite{BEAMnrcManual}. The FLATFILT component module allows for a stacked set of coaxial cones to be specified with independent top (RTOP) and bottom (RBOT) radii \cite{BEAMnrcManual}. By setting equal RTOP and RBOT for all layers, 
The simulation of the scanning system was split into two parts. First the tube housing was simulated and a phase space (phsp) file scoring the energy, position, direction and interaction history of each particle was recorded. The phsp file immediately at the back of the exit window of the x-ray tube (phsp1) was in turn used as the input to the simulation of particle transport through the imaging object. The source-to-object distance was set to 75mm. A second phsp file (phsp2) was placed at 115mm recording particles reaching the detector plane. Our imaging object was designed using the FLATFILT component module to be a uniform water cylinder containing at least one cylindrical layer of contrast material to allow for decomposition of $m \geq 2$ materials. The layer(s) of contrast material(s) and the water cylinder had a radius of 3mm and 6mm around the beam axis, respectively. Material thicknesses were defined in section \ref{ch4_met:val} to be the same as in \cite{NikOptimal}. Spaces at the back of the x-ray tube filtration and between the imaging object and the detector plane consisted of air specified by the SLABS component module. 

Cross sections including Rayleigh scattering were generated from the XCOM dataset using the PEGS4 code system for all the materials used in this work. Directional bremsstrahlung splitting and photon forcing were used in the x-ray production to improve simulation efficiency. The bremsstrahlung splitting field radius and the source-to-surface distance of the splitting field used were 2.8\,cm and 13.5\,cm, respectively. NIST bremsstrahlung cross-section data was used. All Monte Carlo simulations were run with $3\times 10^8$ primary histories and the cut-off energy was 1keV for both electrons and photons. 

One of the main differences between the BEAMnrc simulation and the optimization algorithm described in \cite{NikOptimal} is the inclusion of scattered radiation. In BEAMnrc, the interaction of each particle with the imaging object was tracked via the LATCH bit identification tag to create additional images/spectra with only primary photons. Particle interactions with the air regions were ignored. Information in the phsp files were decoded particle by particle using an in-house developed Matlab code. The data was organized in a stack of two-dimensional matrices containing particles within 1\,keV ranges to allow for retrospective formation of energy-selective images \cite{RuneVirtualsCT}. %RuneReport
The spatial variation in the photon counts was corrected by using an open beam image of 1 to 120\,keV prior to material decomposition. Spectral distribution, given in photon fluence/keV/incident particles of the simulated phase space file was derived using the BEAM Data Processor (BEAMDP) program \cite{BEAMDPum} distributed with BEAMnrc. 

%\subsection{BEAM Data Processor}\label{ch4_met:beamdp}
%The statistical weights scored on each pixel represents the localized probability of photons being detected. The BEAMnrc package comes with a utility program for analyzing the phase space file generated in the simulation . Spectral distribution, given in photon fluence/keV/incident particles of the simulated phase space file was derived using BEAM Data Processor (BEAMDP). In our simulations, the number of incident particles is the number of primary histories. The simulated photon counts per pixel can therefore be computed at any given energy upon processing phsp2 using BEAMDP. Spectra obtained from the phsp2 files were normalised against the fluence output from BEAMDP for the conversion of statistical weights into photon fluence. 

\subsection{Thickness estimation} \label{ch4_decom}
The pixelated measurements were binned as input to $\mathbi{x}$ in (\ref{eq:zscorebins}) for estimation of $\mathbi{t}$. Material decomposition was performed pixel-by-pixel using the spectrum scored in phsp2 in a 128 $\times$ 128 pixel detector grid of \SI{220x220}{\um} each. A direct way to find the solution for (\ref{eq:zscorebins}) is by mapping a look-up table of counts for an extensive sample of thicknesses. The solution can then be provided by locating the thicknesses that are most consistent with the binned measurements:
\begin{equation}
\mathbi{t} = \argmin_{ \mathbi{t} }   { \left\{  \left[  \sum_{k=1}^n  \Big( x_k - \lambda_k(\mathbi{t}) \Big) ^2 \times \frac{1}{ \lambda_k( { \mathbi{t} } ) } \right]  \times \frac{1} {n} \right\}^{\frac{1}{2}} }.
	\label{eq:zscoremin}
\end{equation}
The accuracy of the solution given by the look-up table, however, is dependent on the sample size \cite{AlvarezEstimator} and a huge set of data points may therefore be required for sufficient accuracy. 
%Alvarez \cite{AlvarezEstimator} formulated a maximum likelihood approach by utilizing a calibration phantom to approximate the effective attenuation, which is subjective to the selection of regional values and may require a further error correction computation. 
In this work, a more direct approach was realized by implementing an iterative search algorithm, which implements the Nelder-Mead algorithm \cite{LagariasNelder-Mead}. This was carried out for both the simulated projection data with and without the inclusion of scattered radiation in the BEAMnrc model. By using the look-up table solution as our initial estimates, the Mahalanobis distance in equation \ref{eq:zscorebins} was minimized using the Matlab {\it fminsearch} function without requiring the likelihood function. Furthermore, the determination of the effective attenuation over an energy range can be avoided \cite{NikOptimal}.  

%In their decomposition, Firsching {\it et al.} \cite{FirschingQuantitative} employed the mass attenuation coefficient ($\frac{\mu_{i}}{\rho}$) to reconstruct the areal density $a_{i} := \rho_{i} \cdot t_{i}$ of the materials. Equivalently, $\mu_{i}$ in equation \ref{eq:Nintegral} can be substituted by $\frac{\mu_{i}}{\rho}$ and $a_{i}$ (in place of $t_{i}$) of the materials can be determined when the mass attenuation coefficients are chosen as the basis functions in the proposed method. Likewise, attenuation components of the photoelectric and the Compton effect contributions together with a total mass attenuation of a K-edge material can be incorporated for the determination of basis-material densities along the x-ray path. 

\subsection{Validation of optimal material discrimination}\label{ch4_met:val}

For a constant x-ray tube voltage and current, the theoretical model in \cite{NikOptimal} provided a solution of choosing energy bins for spectral imaging based on the smallest confidence region under the influence of Poisson statistics. To reiterate, a limitation of this model is that it does not take into account scattered radiation. To achieve optimal spectrum weighted attenuation difference in discriminating 0.01\,cm of iodine and 1.5\,cm of water, Nik {\it et al.} \cite{NikOptimal} showed that the optimal bin border ($E_{(h,1)}$) is at \SI{60}{\kilo\electronvolt}. When $E_{(h,1)}$ is fixed at the iodine K-edge of \SI{33}{\kilo\electronvolt}, the optimal higher bin border ($E_{(h,2)}$) was found to be at \SI{51}{\kilo\electronvolt} for the discrimination of iodine, calcium and water. 

Using the BEAMnrc framework, projections for an object consisting of $\tau_{I}$ = 0.01\,cm of iodine between two 0.75\,cm cylindrical layers of water background ($\tau_{H_2O}$ = 1.5\,cm) were simulated. 
%The discrimination of more materials using the reconstructed image has been proposed via mass/volume conservation \cite{LiuDECT3materials,PaulQuantifying} or by segmenting pixels into classes of materials prior to the decomposition \cite{LeSegmentation,AlessioQuantitative}. In the projection space, the same may be achievable by assigning up to 4 dimensionality to the linear attenuation coefficients \cite{BornefalkXCOM}. Here, we focus on the decomposition of up to 3 materials. 
To decompose 3 materials, the projection data of $\tau_{I}$ = 0.01\,cm and $\tau_{Ca}$ = 0.22\,cm stacked between two 0.75\,cm cylindrical layers of water background was simulated. The density for iodine and calcium was defined to be the same as in \cite{NikOptimal}, i.e. \SI{4.93}{\gram\per\cubic\centi\metre} and \SI{1.55}{\gram\per\cubic\centi\metre}, respectively. 

For a given incident x-ray spectrum, a pertinent problem is to determine the minimum exposure to achieve an imaging task. The Rose's criterion \cite{RoseCriterion} of SNR $\geq$ 5 is often used as a target for image quality (e.g. in \cite{DucoteNanoparticles}). 
When decomposing a homogenous material {\it i} with thickness $\tau_i$, the SNR within the uniform region-of-interest (ROI) can be provided by the ratio of the reference thickness to the standard deviation of thickness population, $(\tau_i / \sigma_i)$. Likewise, in estimating the material quantity in a pixel, $\sigma_i$ represents the uncertainty in the estimation. An imaging task can thus be setup as achieving the $\tau_i / \sigma_i$ value of 5, in the quantification of thickness $\tau_i$, or in the homogenous ROI of the decomposed image {\it i}. The minimum number of photons per unit area required in order to accomplish the imaging task can be subsequently computed to fulfill the ALARA principle \cite{SlovisALARA}. 

To directly compare with the BEAMnrc MC simulation in this work, however, the image noise was estimated for the simulated detected counts. Using the theoretical model, the image noise was computed as variance $(\sigma_i^2)$ as in \cite{RoesslSensitivity} and \cite{DucoteNanoparticles}. The diagonal elements of the covariance matrix described in section \ref{background} incorporates $\sigma_i^2$ and can therefore be utilized for the prediction of image noise (or SNR). This enables a direct comparison between the $\sigma_i^2$ values obtained from the metric and the simulation. For the discrimination of iodine/water, $\sigma^2$ was determined at an interval of \SI{1}{\kilo\electronvolt} for $E_{(h,1)}$ ranging from \SIrange{20}{100}{\kilo\electronvolt}, whereas $E_{(h,1)}$ was fixed at \SI{33}{\kilo\electronvolt} and $\sigma^2$ was computed for $E_{(h,2)}$ between \SIrange{36}{100}{\kilo\electronvolt} for the discrimination of iodine, calcium and water. 

In the BEAMnrc model, the precision of material decomposition was examined by determining the image noise 
%(variance = $\sigma^2$, where $\sigma$ is the standard deviation) 
of the material basis images. Mean and variance were computed for the central 690 pixels in the region with contrast material(s). The simulated variance was computed for bin border energies ranging from \SIrange{20}{100}{\kilo\electronvolt} for the decomposition of two materials and \SIrange{36}{100}{\kilo\electronvolt} for the decomposition of three materials, as in the theoretical model. Bin border energies below \SI{20}{\kilo\electronvolt} and above \SI{100}{\kilo\electronvolt} were considered suboptimal in both models due to photon starvation. 

While variance is given by the averaged difference between the thickness output and its mean thickness value, another important measure for material quantification is the averaged difference between the output and the actual value of thicknesses, known as the bias. The mean square error (MSE) incorporates both the bias and variance. The following figure of merit (FOM) was therefore formulated as a validation of the theoretical model in \cite{NikOptimal}:
\begin{equation}
	\textup{FOM}= \left( { \sum_{i=1}^m MSE_i/\tau_i^2} \right)^{-\frac{1}{2}}. 
	\label{eq:fomMSE}
\end{equation}
	
(\ref{eq:fomMSE}) was evaluated for bin border energies ($E_{(h,1)}$) from \SIrange{20}{100}{\kilo\electronvolt} for the decomposition of iodine and water. 
%To investigate on the effect of decomposing two materials using three energy bins, the variance and MSE were determined with the lower bin border energy ($E_{(h,1)
For the decomposition of 3 materials, the lower bin border energy ($E_{(h,1)}$) was held at the K-edge of iodine (\SI{33}{\kilo\electronvolt}), while a FOM curve was plotted for the upper bin border energies ($E_{(h,2)}$) ranging from \SIrange{36}{100}{\kilo\electronvolt} for the higher energy bin to validate the results in \cite{NikOptimal}.

\section{Results}\label{results}

%\subsection{Monte Carlo simulations}\label{res:MCsim}
%\begin{figure}[tbp]
%  \centering
%   \includegraphics[width=0.25\textwidth, angle=-90]{fig_BEAMvsSpecgen.eps}
%    \caption{Comparison of the simulated incident energy spectrum with a reference spectrum from Specgen (Glenn Stirling, NRL, Christchurch, New Zealand) obtained using Tucker {\it et al.}'s \cite{TuckerW} model of W/Rh target shows negligible differences. } 
%   \label{fig_BEAMvsSpecgen}
%\end{figure} 

%The 120\,keV monoenergetic electrons incident on the x-ray anode described in section \ref{ch4_met:MCsim} produced approximately $1.2\cdot10^8$ photons. The open beam spectrum on the detector plane was used for the aforementioned normalization of incident counts in the optimization algorithm and the conversion of statistical weights into photon fluence. Figure \ref{fig_BEAMvsSpecgen} demonstrated negligible differences between the MC calculated spectrum and an analytical calculation using the Tucker {\it et al.}'s \cite{TuckerW} model generated by Specgen (Glenn Stirling, National Radiation Laboratory, Christchurch, New Zealand). The uncertainty of the simulation given by BEAMDP was plotted as error bars on the spectrum. A comparison of Specgen with other packages such as xcomp5r and TASMIP can be found in \cite{MeyerSpecgen}. 

\subsection{Validation of optimal material discrimination}\label{ch4_res:val}

\begin{figure}[tbp]
\centerline{ \subfigure[] {\includegraphics[width=0.28\textwidth, angle=-90]{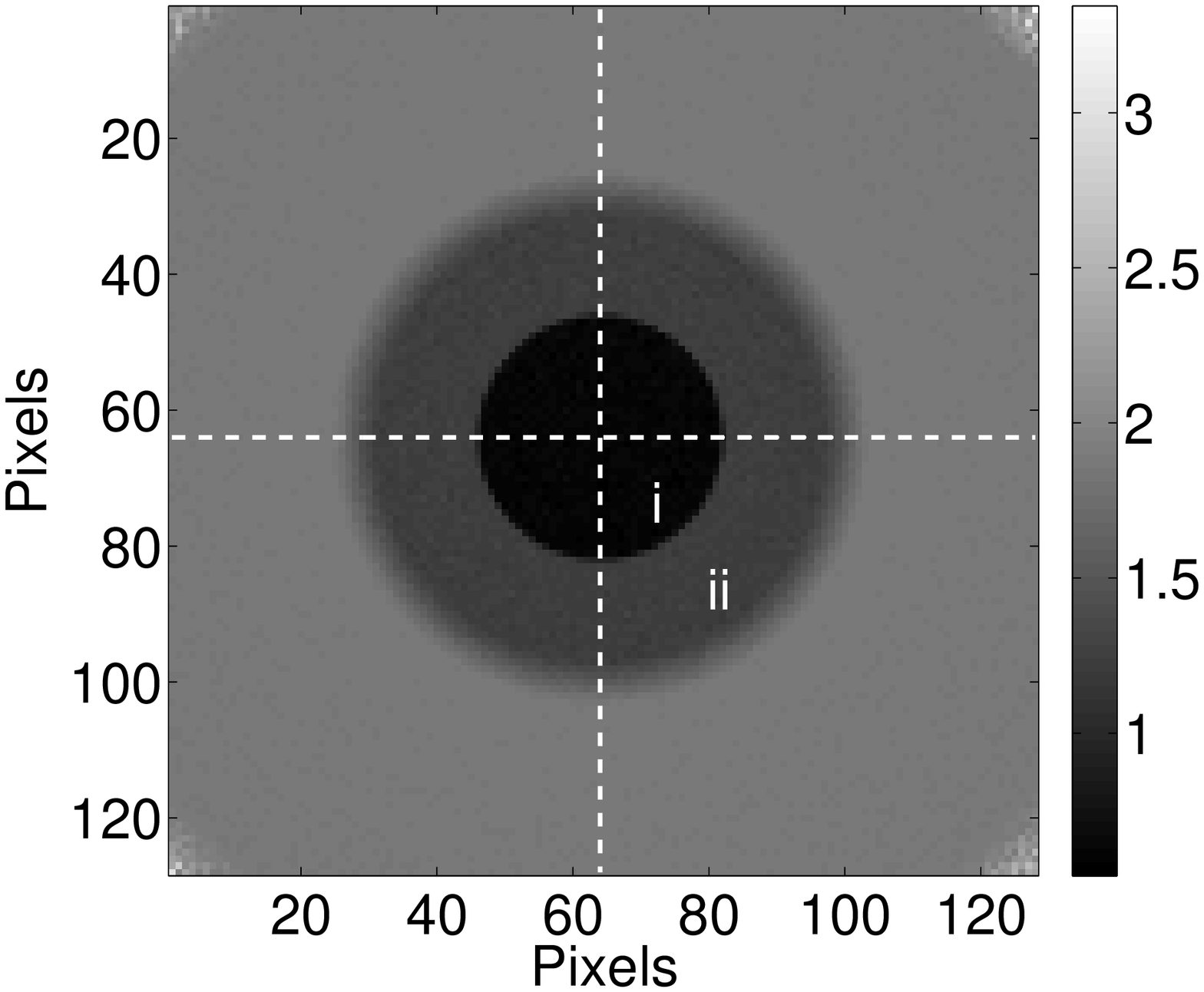}%
\label{fig_sampleInProj}}
\hfil
\subfigure[]{ \includegraphics[width=0.28\textwidth, angle=-90]{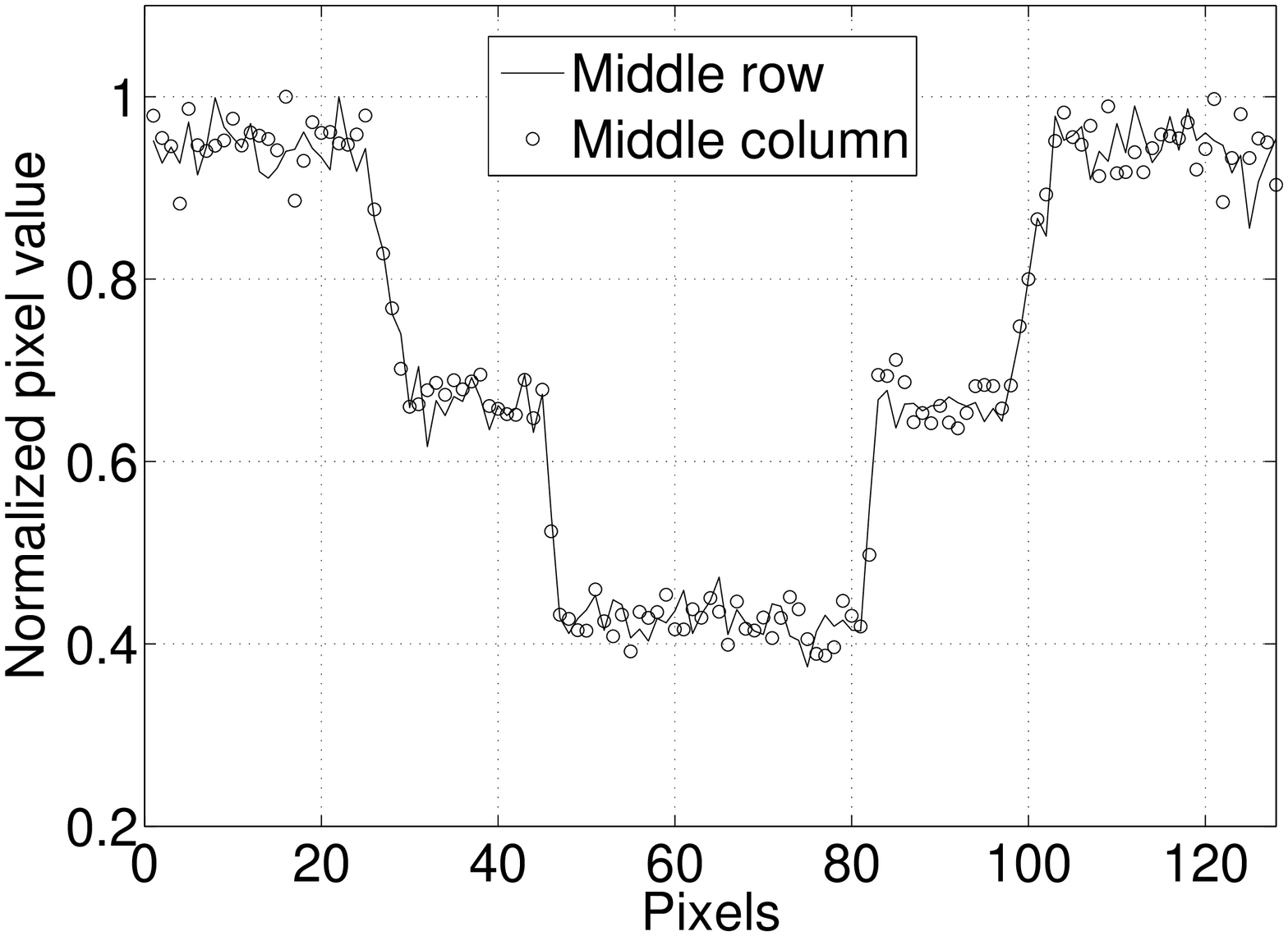} %
\label{fig_sampleInProfile}}}
\caption{A representative (a) projection image and (b) set of profiles upon normalization using the open beam image. Color bar in (a) indicates an arbitrary unit upon normalization. The inner (region i) and outer (region ii) concentric circular regions are the ROIs with and without contrast material(s) within the water cylinder, respectively. Statistical analysis was performed on the pixels within the region i. 
(b) The profile across the horizontal axis (solid line) is relatively constant and is used as a reference to show no reminiscence of the Heel effect in the corrected middle column profile (circles) after normalization against the open beam.} 
\label{fig_OpenBeamCorrection}
\end{figure}

A representative set of projection images in figure \ref{fig_sampleInProj} shows two concentric circular regions. The darker inner region (i) shows the pixels with higher attenuation due to the contrast material(s) within the water cylinder and the outer mid-gray region (ii) represents the water region without contrast material. While decomposition was performed on the full-field projections, only the ROI with the overlapping contrast materials (region i) was analyzed. 

\begin{figure}[tbp]
\centerline{
\subfigure[] {\includegraphics[width=0.24\textwidth, angle=-90]{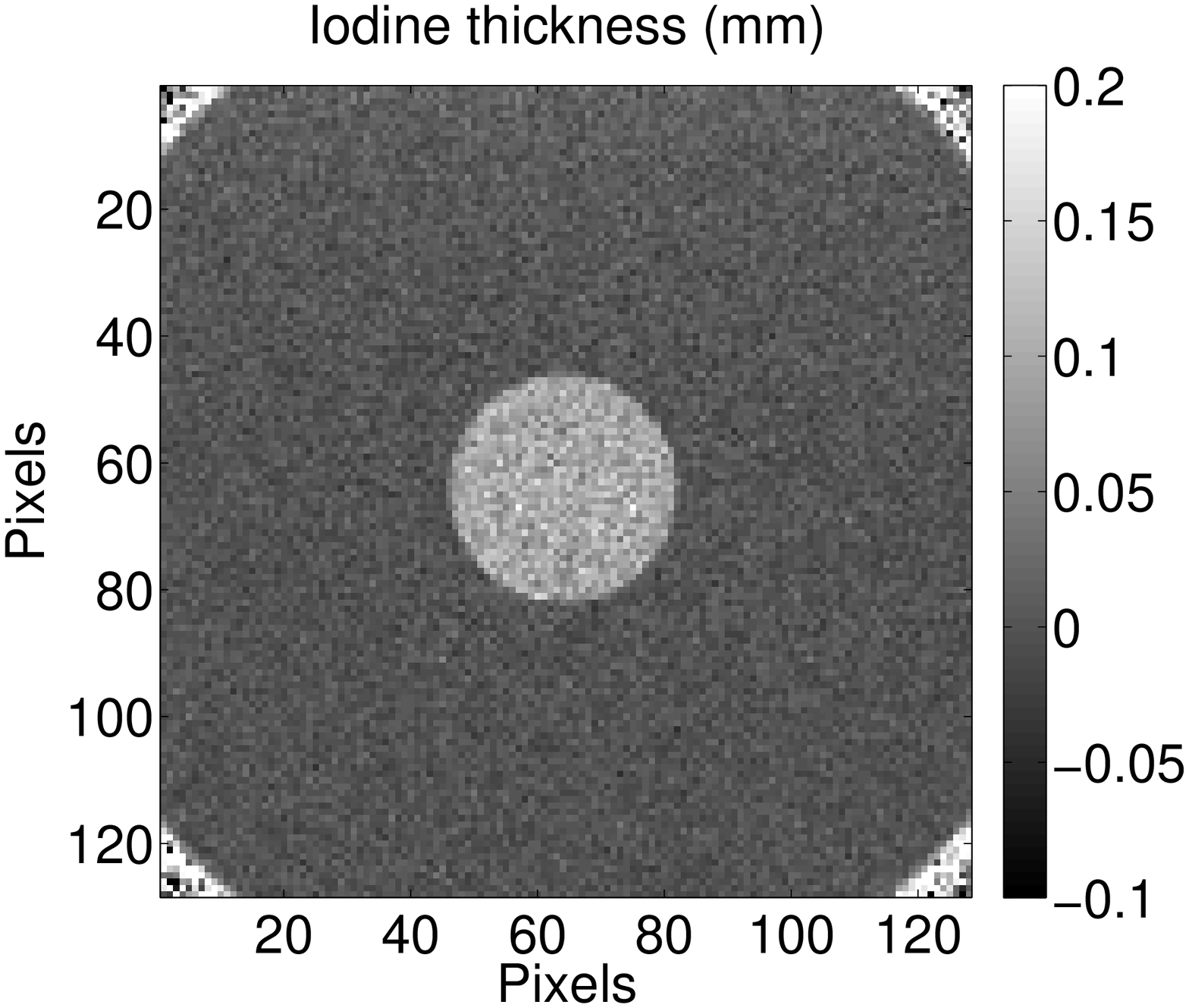}%
\label{fig_IbasisIm}} 
\subfigure[]{ \includegraphics[width=0.24\textwidth, angle=-90]{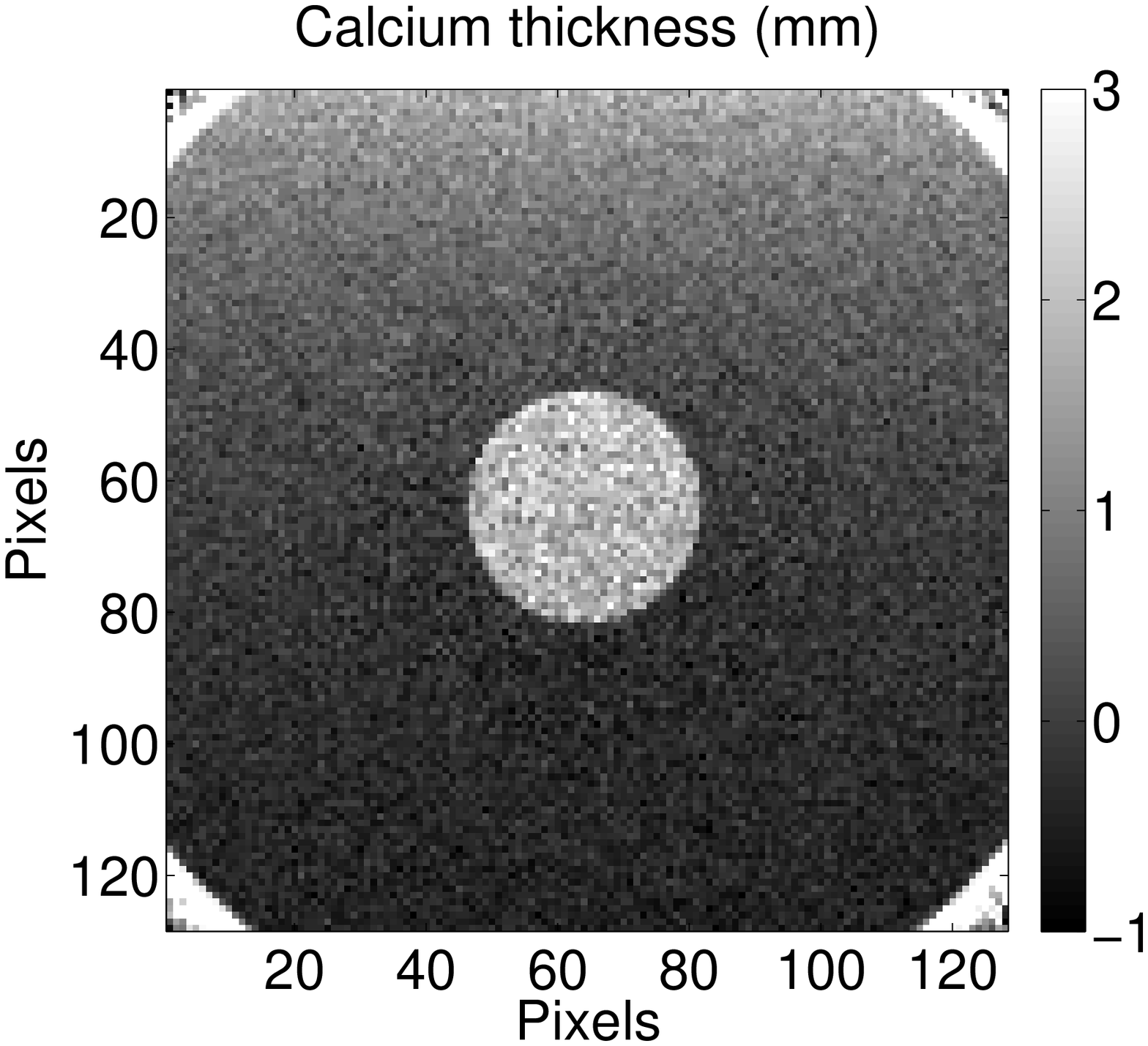}  %
\label{fig_CabasisIm}} 
\subfigure[]{ \includegraphics[width=0.24\textwidth, angle=-90]{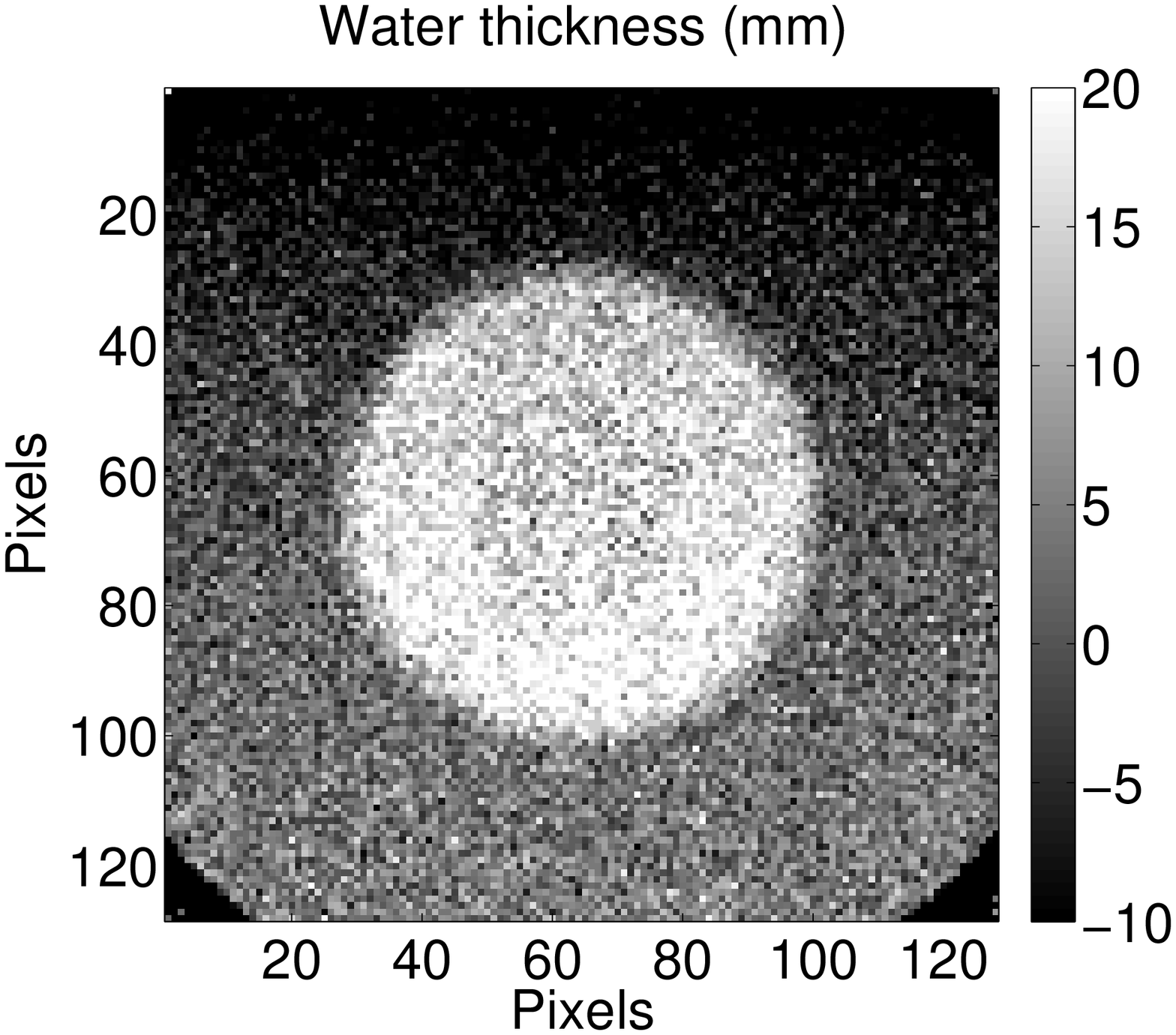}  %
\label{fig_H2ObasisIm}} }
%\centerline{
%\subfloat[] {\includegraphics[width=0.25\textwidth, angle=-90]{fig_Iprofile}%
%\label{fig_Iprofile}} 
%\subfloat[]{ \includegraphics[width=0.25\textwidth, angle=-90]{fig_Caprofile}  %
%\label{fig_Caprofile}} 
%\subfloat[]{ \includegraphics[width=0.25\textwidth, angle=-90]{fig_H2Oprofile}  %
%\label{fig_H2Oprofile}} }
%The respective profiles (d - f) through the middle row of pixels were normalised to $\tau_{i}$ to illustrate the accuracy of the estimations. 
\caption{ The decomposed material basis images of (a) iodine, (b) calcium and (c) water. A  quantitative analysis of the accuracy as well as the precision can be found in figure \protect\ref{fig_imageNoise2mat}, figure \protect\ref{fig_imageNoise3mat} and table \protect\ref{tab:MSE}. }
\label{fig_basisIm}
\end{figure}

\begin{figure}[tbp]
\centerline{ \subfigure[] {\includegraphics[width=0.28\textwidth, angle=-90]{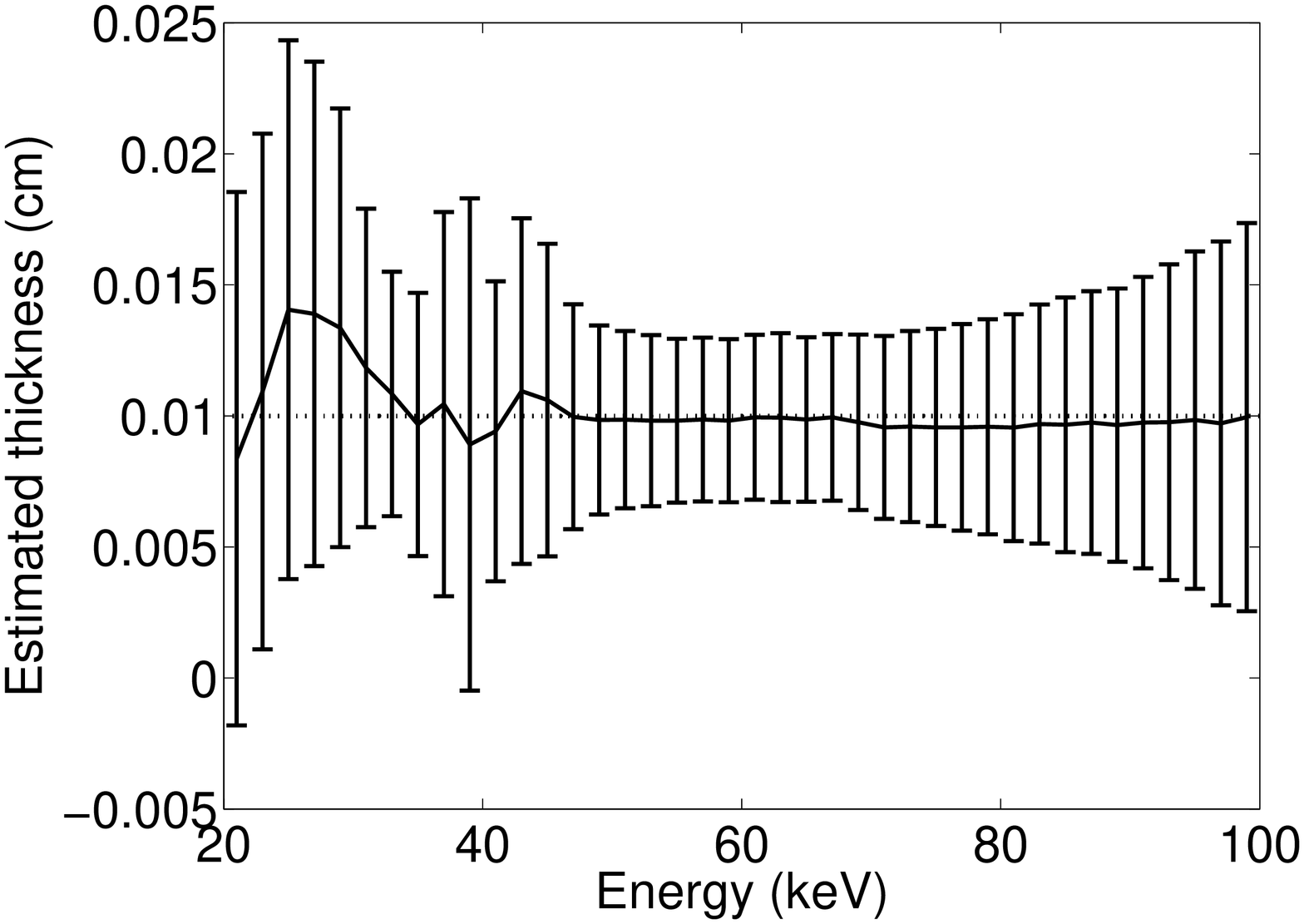}%
\label{fig_terrorbar2matI}}
\hfil
\subfigure[]{ \includegraphics[width=0.28\textwidth, angle=-90]{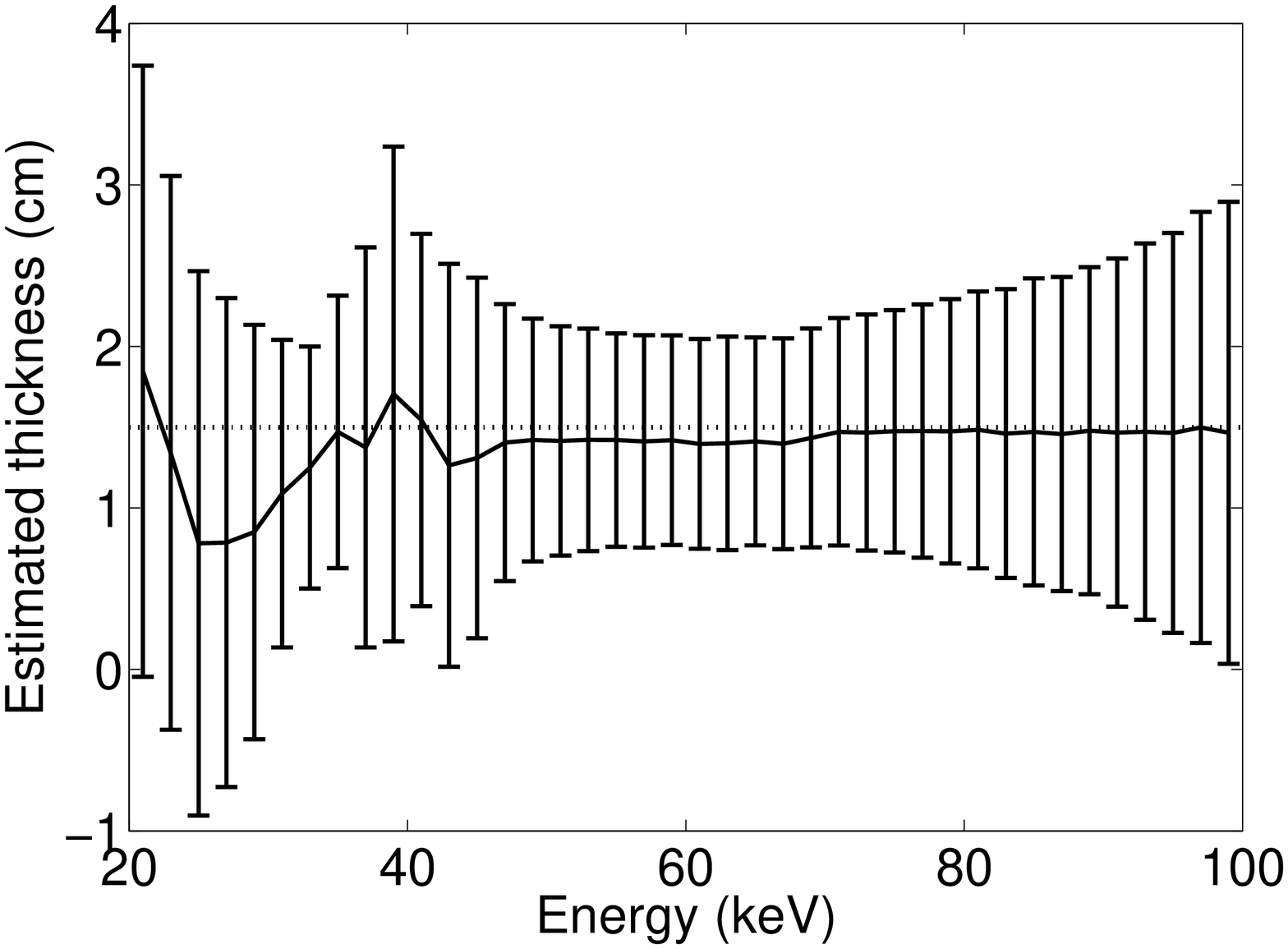} %
\label{fig_terrorbar2matH2O}}}
\caption{Mean and standard deviation obtained for the material basis images for (a) iodine and (b) water, displayed at an interval of \SI{2}{\kilo\electronvolt}. The calculation of these error bars allows the quantification of image noise and comparison with the theoretical prediction %(figure \ref{fig_imageNoise2mat}), 
which leads to validation the optimization of bin border energy based on our FOM. Optimal bin border energies are indicated by the smallest error bars. The average bias over \SI{5}{\kilo\electronvolt} around the optimal bin border is 1.24\% for (a) and 6.06\% for (b).}
\label{fig_errobar2mat}
\end{figure}

\begin{figure}[tbp]
\centerline{ \subfigure[] {\includegraphics[width=0.24\textwidth, angle=-90]{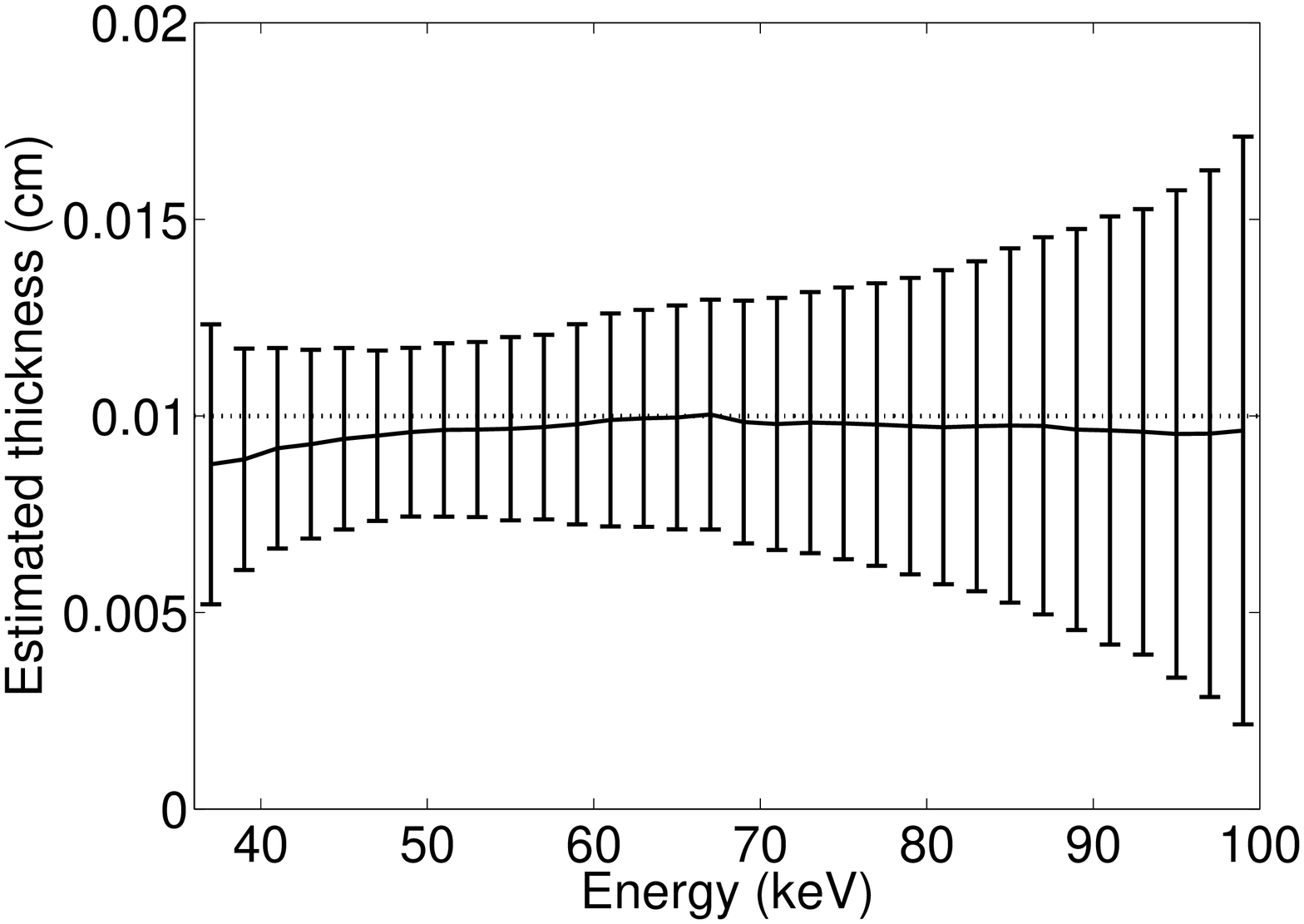}%
\label{fig_terrorbar3matI}}
\hfil
\subfigure[]{ \includegraphics[width=0.24\textwidth, angle=-90]{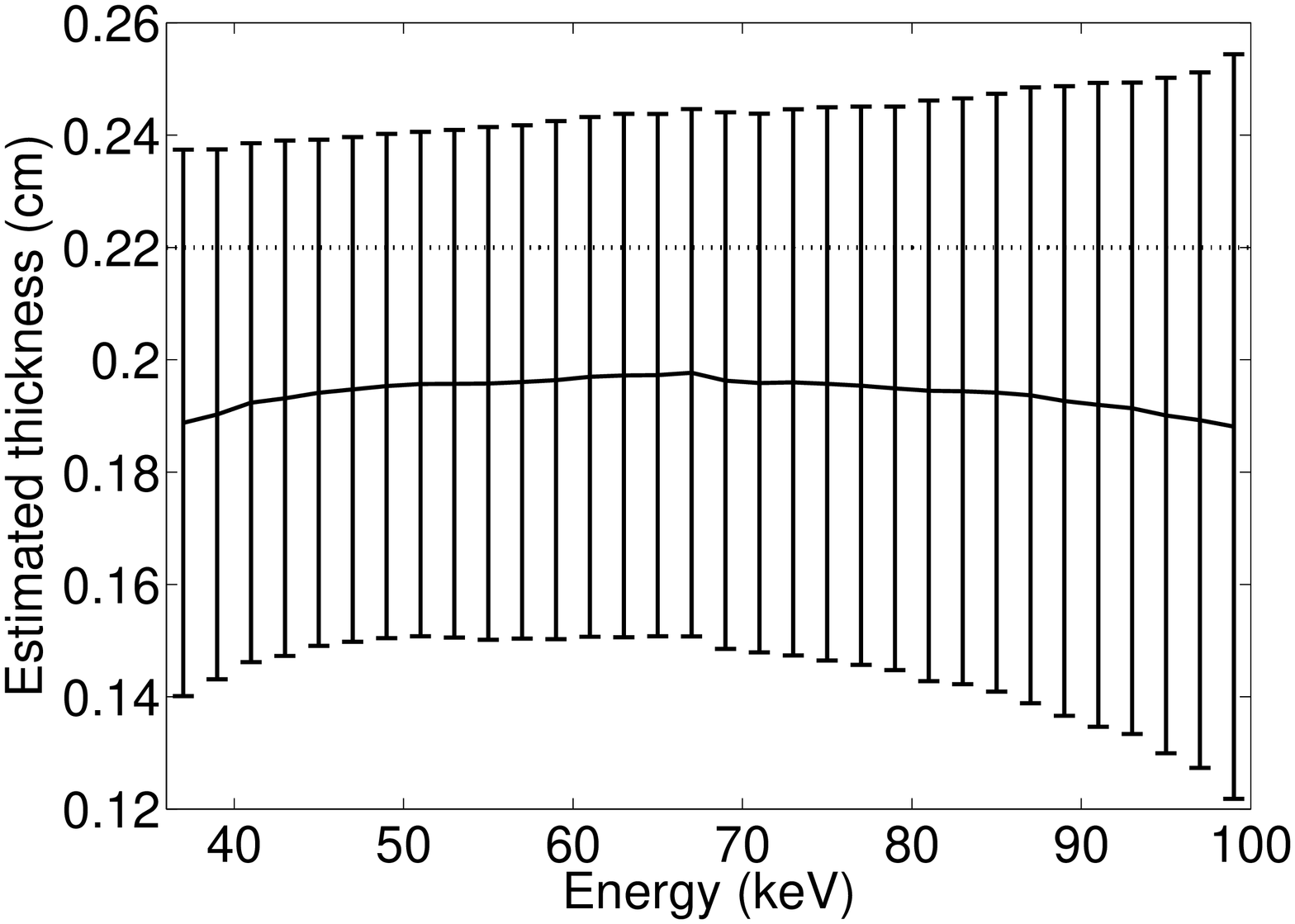} %
\label{fig_terrorbar3matCa}}
\hfil
\subfigure[]{ \includegraphics[width=0.24\textwidth, angle=-90]{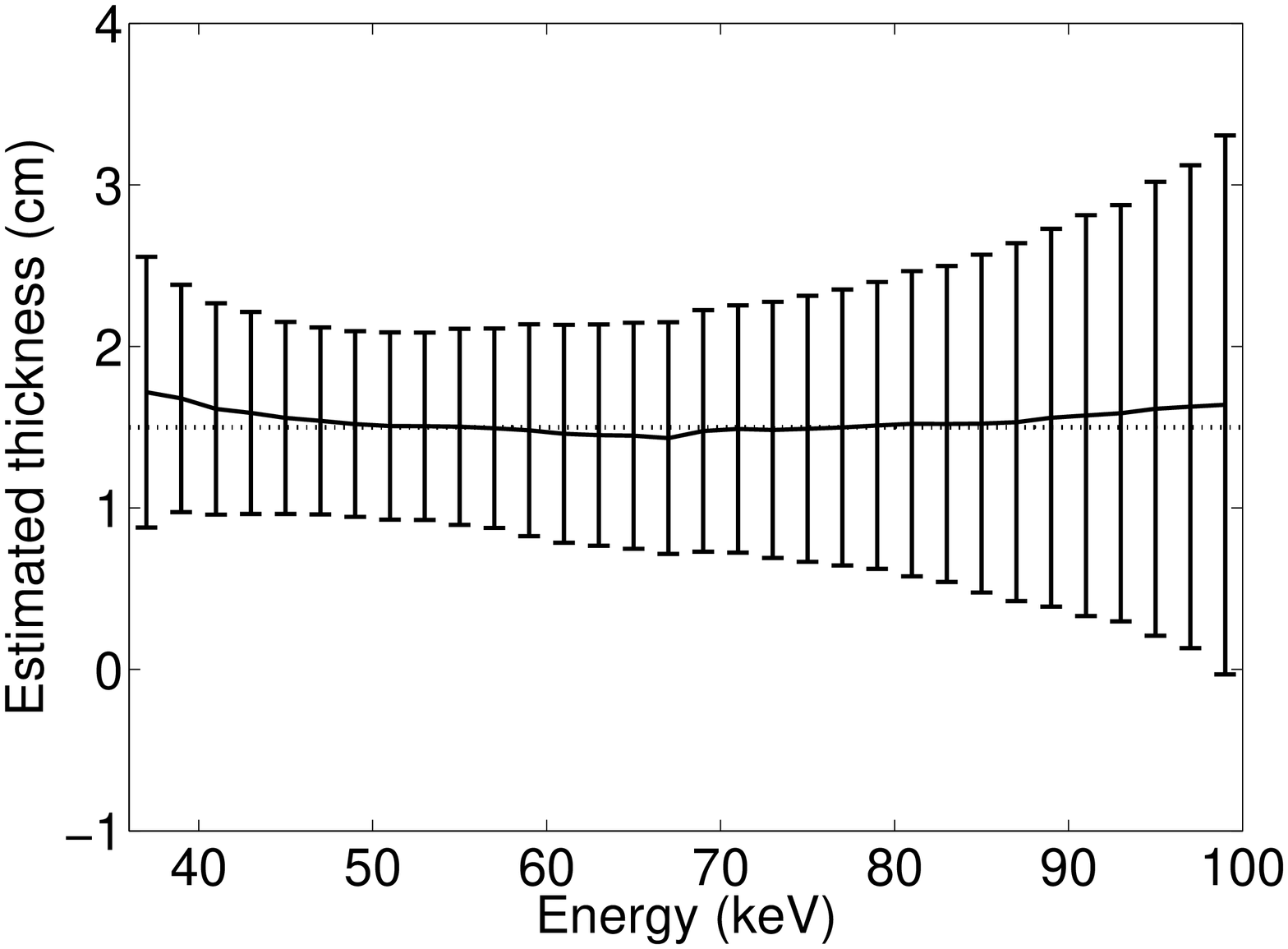} %
\label{fig_terrorbar3matH2O}}}
\hfil
\centerline{ \subfigure[] {\includegraphics[width=0.24\textwidth, angle=-90]{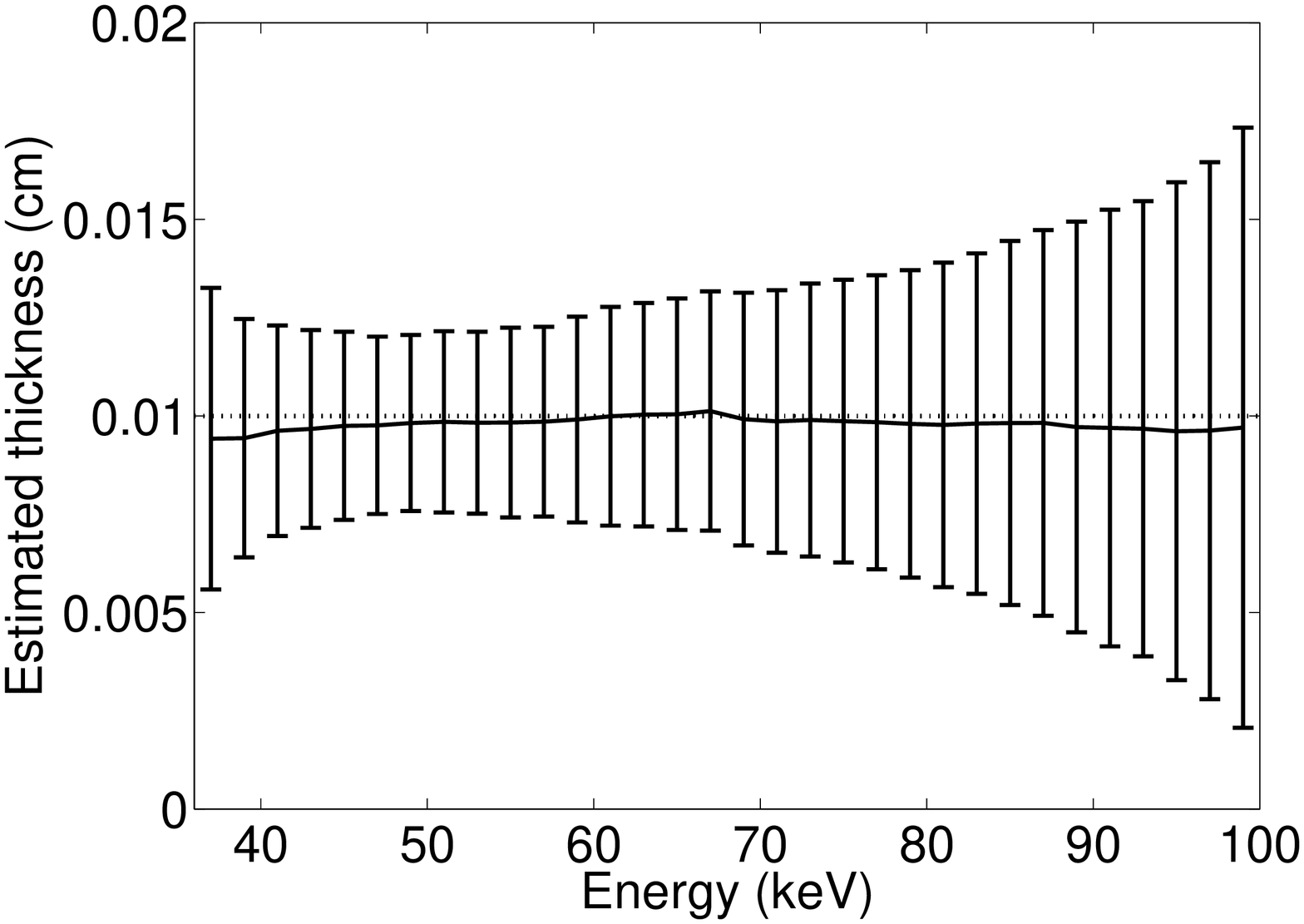}%
\label{fig_terrorbar3matISR}}
\hfil
\subfigure[]{ \includegraphics[width=0.24\textwidth, angle=-90]{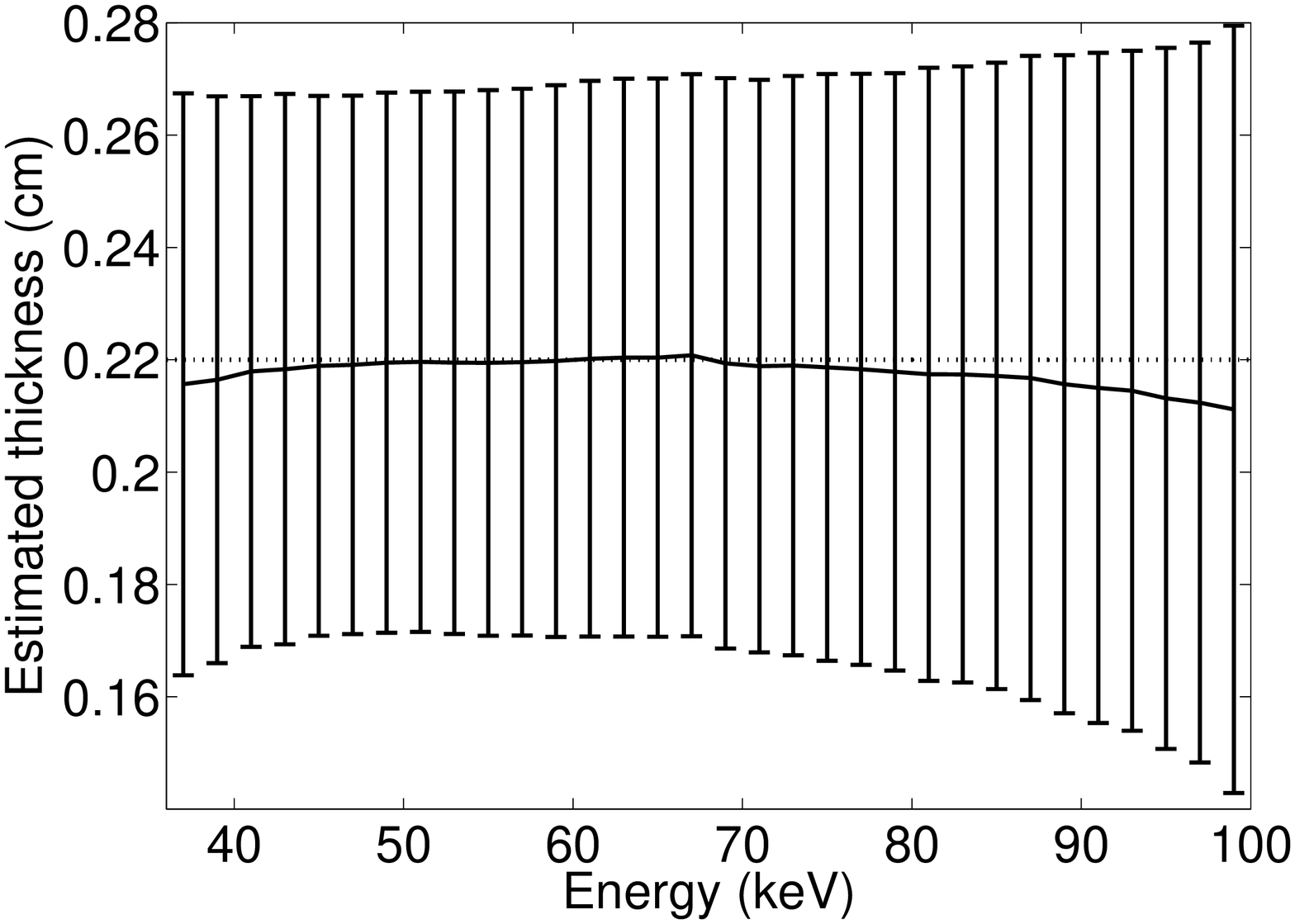} %
\label{fig_terrorbar3matCaSR}}
\hfil
\subfigure[]{ \includegraphics[width=0.24\textwidth, angle=-90]{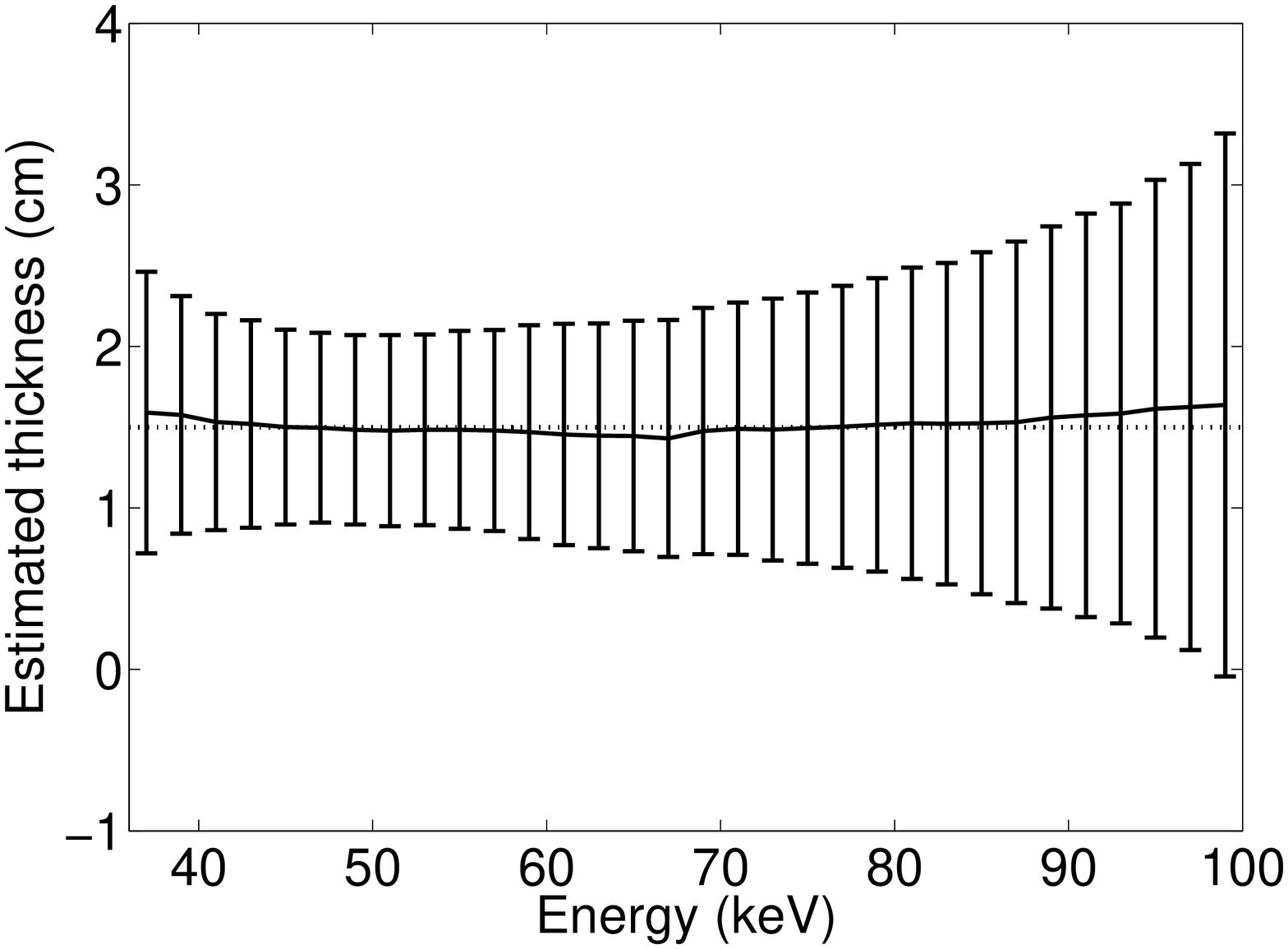} %
\label{fig_terrorbar3matH2OSR}}}
\caption{ Mean and standard deviation obtained for the material basis images for (a,d) iodine (b,e) calcium and (c,f) water, displayed at an interval of \SI{2}{\kilo\electronvolt}s. The average bias over \SI{5}{\kilo\electronvolt} around the optimal bin border is 3.88\% for (a), 11.14\% for (b) and 0.91\% for (c). The improvement on the bias upon the rejection of scattered radiation is shown in the lower panel. Rejection of scattered radiation improved the accuracy of material quantification, particularly in the calcium basis image (see table \protect\ref{tab:MSE}). A comparison with the image noise predicted by the theoretical model is shown in figure \protect\ref{fig_imageNoise3mat} .}
\label{fig_errobar3mat}
\end{figure}

%, with their respective profiles through the middle horizontal axis plotted in the bottom row of figure \ref{fig_basisIm} The profiles were normalized with respect to $\tau_{i}$, which produced values around unity in the presence of materials, demonstrating the accuracy of the decomposition. 
Figure \ref{fig_basisIm} shows a representative set of material basis projection images decomposed using equation \ref{eq:zscoremin}. A quantitative measurement of the decomposition's precision and accuracy is summarized in figures \ref{fig_errobar2mat} and \ref{fig_errobar3mat}. The solid line represents the mean thickness over the 690 pixels within region (ii) in figure \ref{fig_sampleInProj}, whereas the error bars show the standard deviation ($\sigma$) for the decomposition using a particular bin border energy. The reference thicknesses ($\tau_{i}$) was plotted with dotted lines to provide an indication on the bias of the decomposition.

The variance ($\sigma^2$) and the MSE are tabulated in table \ref{tab:MSE} to show the consistency with the estimated image noise given by the theoretical model described in section \ref{background}. Specifically, the theoretical variance (variance$_{A}$), the simulated variance (variance$_{B}$) and the MSE were averaged over the \SI{5}{\kilo\electronvolt} around the theoretical optimal bin border energy, i.e. optimal $E_{(h,1)} \pm $\SI{2}{\kilo\electronvolt} and optimal $E_{(h,2)} \pm $\SI{2}{\kilo\electronvolt} for the decomposition of two and three materials, respectively. The minimal bias around the optimal bin border was reflected in the similar MSE and variance values for the decomposition of two materials. Note that some bin border energy, e.g. \SI{28}{\kilo\electronvolt} for the decomposition of iodine/water in figure \ref{fig_errobar2mat}, provided inaccurate material thicknesses (see section \ref{discussion}). 
%Variances and MSE were both reduced when a third bin was added and optimized as described in section \ref{ch4_met:val}. 

\begin{figure}[tbp]
\centerline{ \subfigure[] {\includegraphics[width=0.28\textwidth, angle=-90]{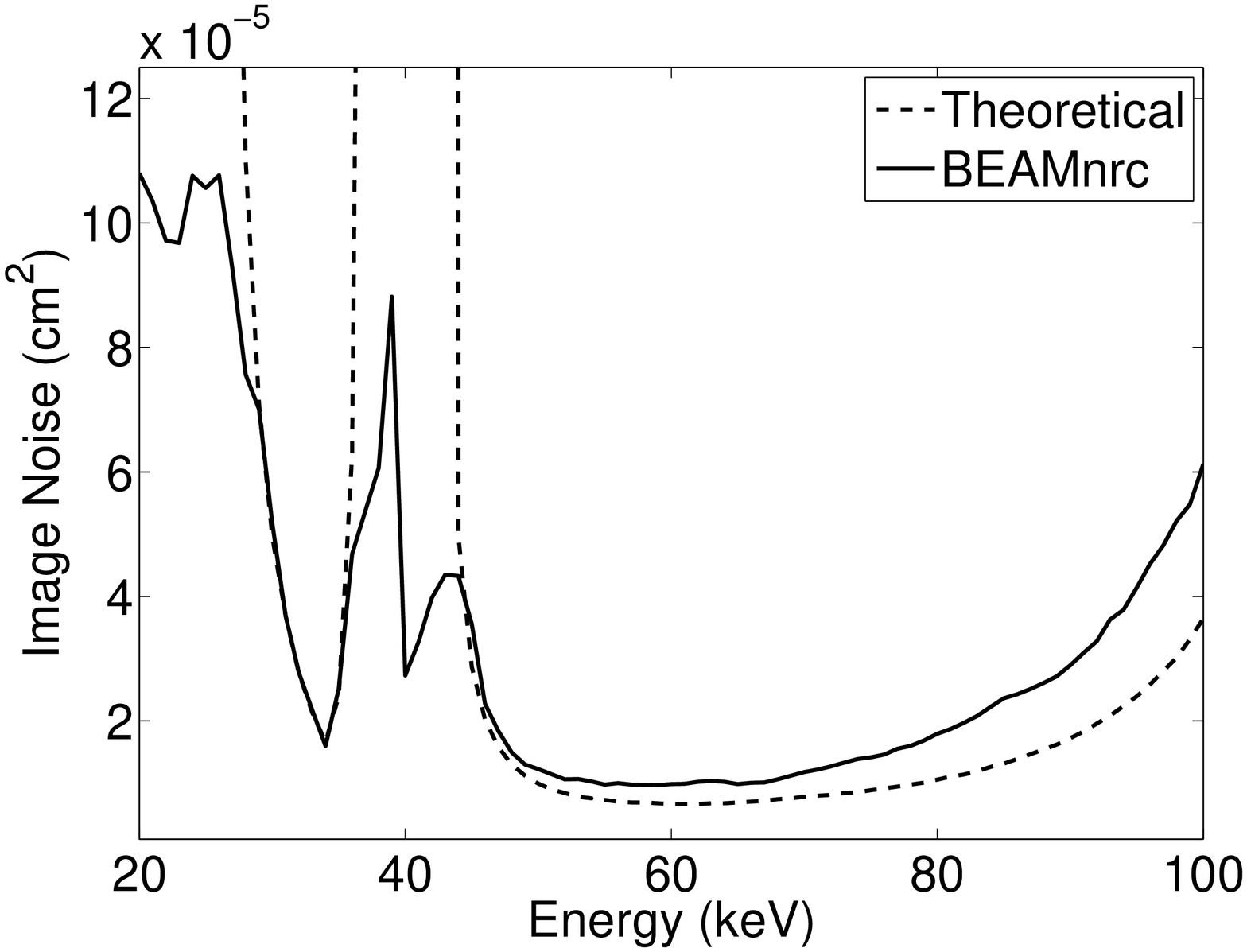}%
\label{fig_imageNoiseI}}
\hfil
\subfigure[]{ \includegraphics[width=0.28\textwidth, angle=-90]{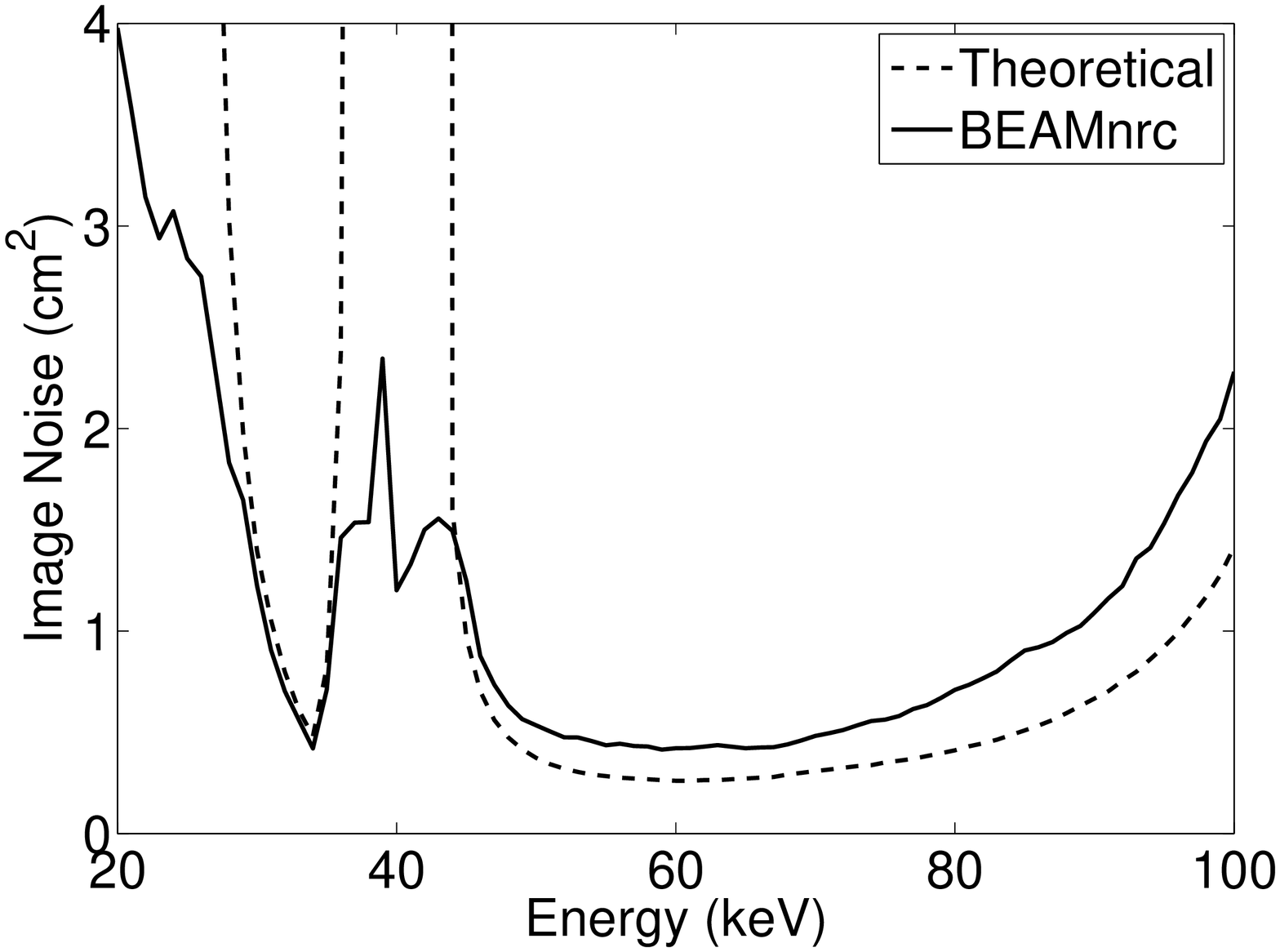} %
\label{fig_imageNoiseH2O}}}
\caption{Image noise of the material basis images as a function of $E_{(h,1)}$ for (a) iodine and (b) water. $\sigma^2$ values obtained from the theoretical model were plotted using dotted lines to show the consistency with the simulated image noise. The theoretical values were allowed to extend beyond the vertical axis to focus on the lowest $\sigma^2$ values. An explanation of this effect will be provided in section \protect\ref{discussion}. A minimization of the $\sigma^2$ values leads to the maximization of FOM and thus the optimization of energy bins.}
\label{fig_imageNoise2mat}
\end{figure}

\begin{figure}[tbp]
\centerline{ \subfigure[] {\includegraphics[width=0.24\textwidth, angle=-90]{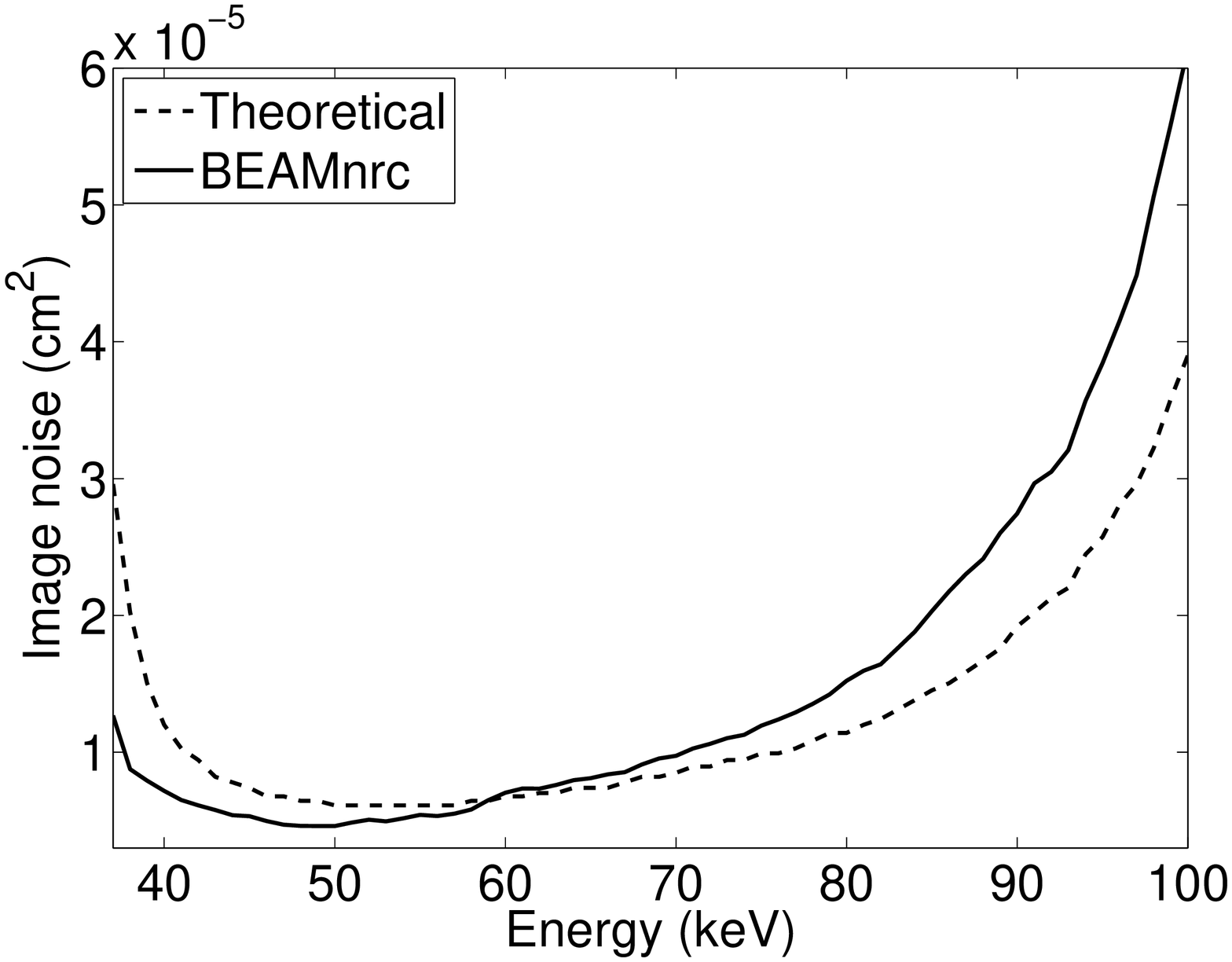}%
\label{fig_imageNoiseI3mat}}
\hfil
\subfigure[]{ \includegraphics[width=0.24\textwidth, angle=-90]{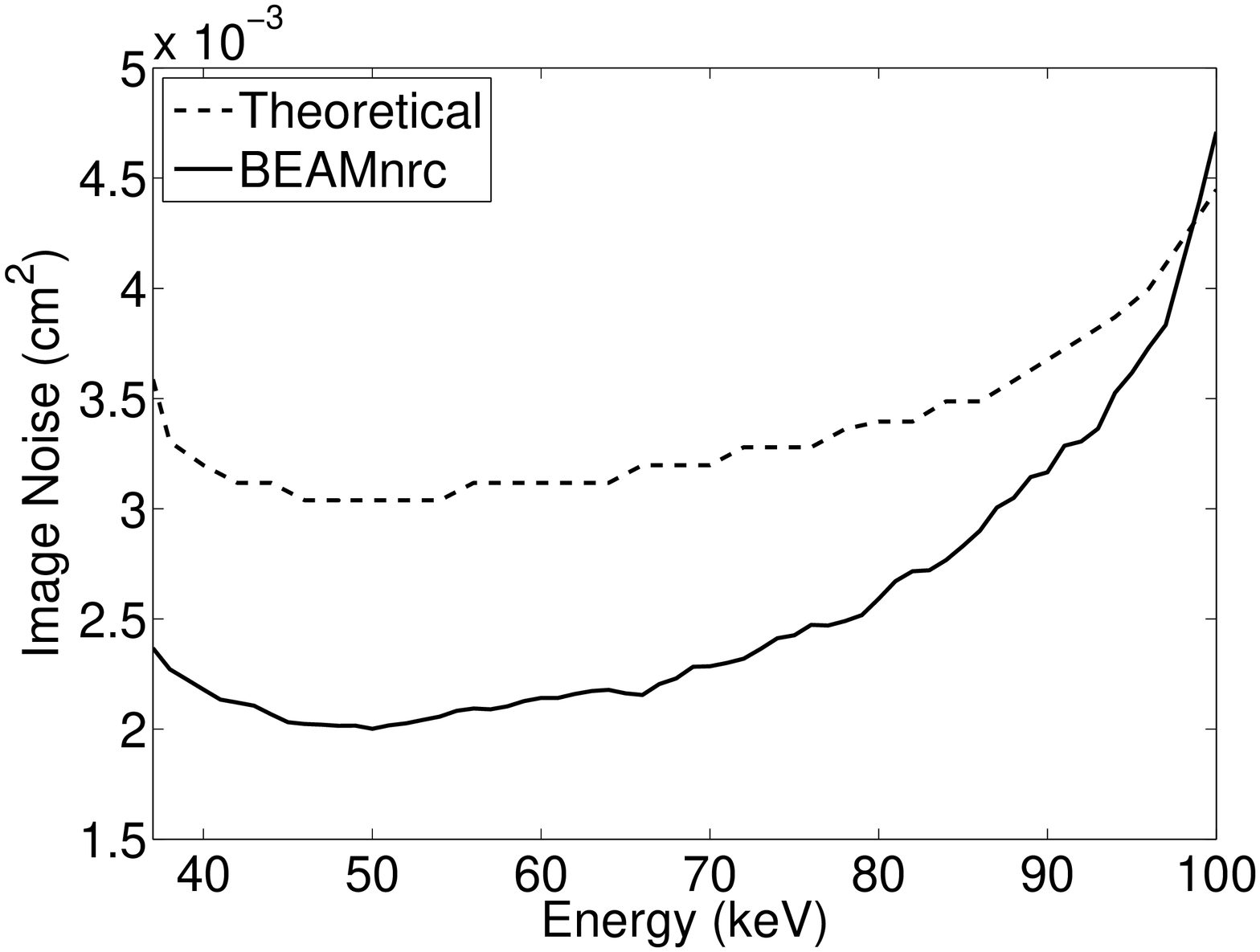} %
\label{fig_imageNoiseCa3mat}}
\hfil
\subfigure[]{ \includegraphics[width=0.24\textwidth, angle=-90]{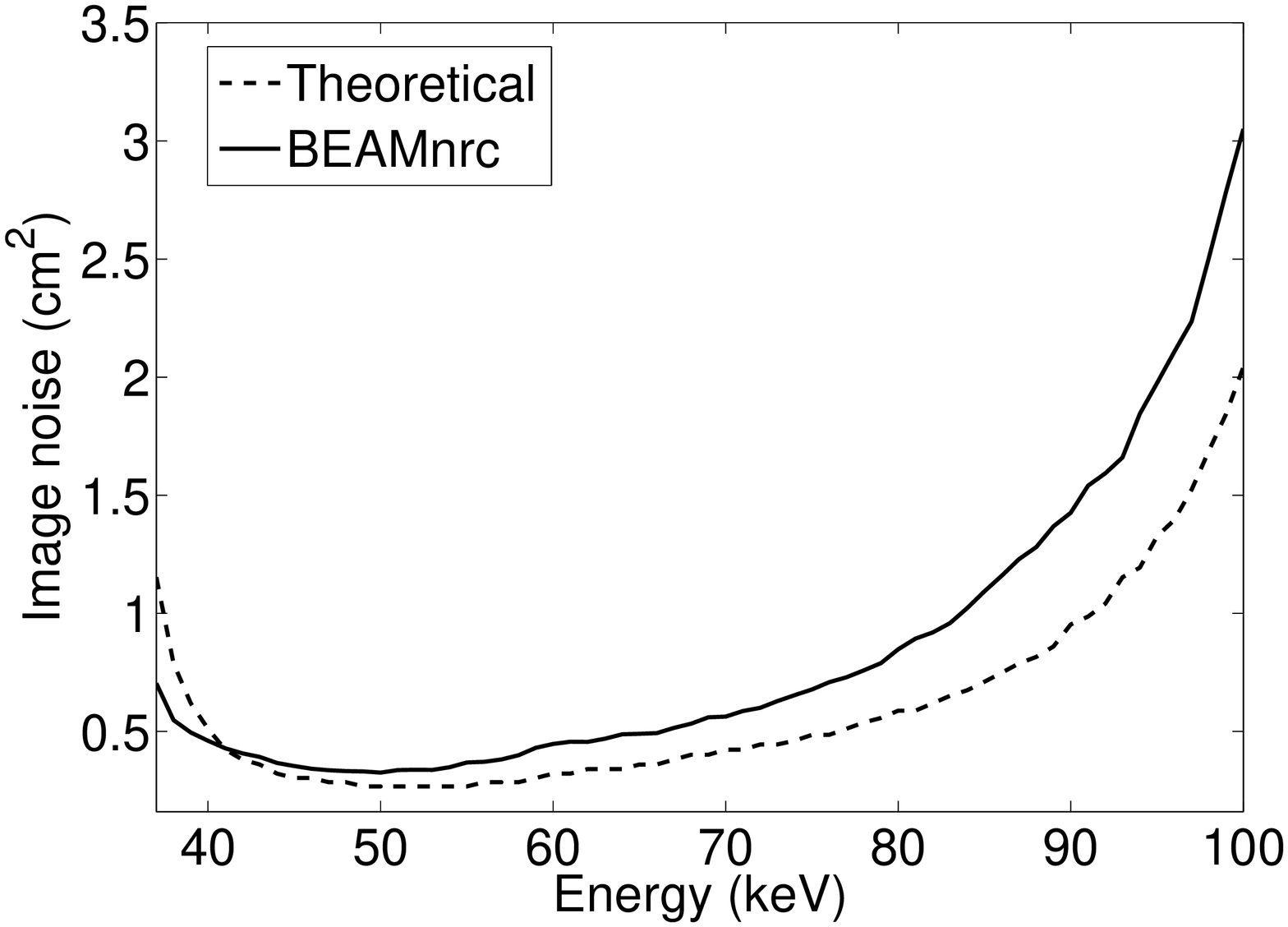} %
\label{fig_imageNoiseH2O3mat}}}
\caption{Image noise of the (a) iodine, (b) calcium and (c) water basis images as a function of $E_{(h,2)}$.}
\label{fig_imageNoise3mat}
\end{figure}

For the case of three materials (figure \ref{fig_errobar3mat}), a higher MSE compared to the variance, particularly for the calcium image, was obtained without the rejection of scattered radiation. This can be seen in the deviation of the solid line from $\tau_{Ca}$ in figure \ref{fig_terrorbar3matCa}. Figure \ref{fig_terrorbar3matCaSR} shows a considerable reduction in the bias of thickness estimation for calcium upon rejection of scattered radiation. Variance$_{B}$, MSE and bias for the decomposition of 3 materials before and after scatter rejection can also be found in table \ref{tab:MSE}. Figure \ref{fig_imageNoise2mat} and figure \ref{fig_imageNoise3mat} show a comparison variance$_{B}$ to variance$_{A}$ for the decomposition of two and three materials, respectively. 
The minimization of the combined $\sigma^2$ in the decomposition leads to the optimization of energy bins. 
%The predicted image noise from the theoretical model were produced using dotted lines to compare with the simulated variances (solid lines). 

%\figuremacroSIII{fig_imageNoise3mat}{Image noise for the decomposition of 3 materials.}{fig_imageNoiseI3mat}{fig_imageNoiseCa3mat}{fig_imageNoiseH2O3mat}{0.225}{-90}{Image noise of the material basis images for (a) iodine, (b) calcium and (c) water.}

%Discrepancy in (b) was due to the memory limit in Matlab, which resulted in a step-wise function for the theoretical prediction of calcium image noise.}

\begin{table*}[tbp]
%\renewcommand{\arraystretch}{1.3}
% if using array.sty, it might be a good idea to tweak the value of
% \extrarowheight as needed to properly center the text within the cells
\centering{
\caption[A summary of mean square error, variance and bias obtained using the theoretical and the BEAMnrc models.]{A summary of mean square error (MSE), variance and bias obtained using the theoretical ({\it A}) and BEAMnrc ({\it B}) models.}

\begin{tabular}{| c | c | c | c | c |}

	\hline
	& 	&  \multicolumn{3}{|c|}{Materials}  \\
 \cline{3-5}
	& & I  (0.01\,cm) & $\mathrm{H_2O}$ (1.5\,cm) & Ca (0.22\,cm) \\
	\hline	
	\multirow{2}{*}{2 materials}	 & variance$_A(\si{cm^2})$	& $6.71 \times 10^{-6}$  &  $2.62 \times 10^{-1}$ & - \\
\multirow{2}{*}{(2 bins)}  &  variance$_B(\si{cm^2})$ & $9.78 \times 10^{-6}$  &  $4.25\times 10^{-1}$ & -\\
	& {MSE $(\si{cm^2})$} & $9.78 \times 10^{-6}$ & $4.33 \times 10^{-1}$ &- \\ 
	
	\hline
%	
%\multirow{2}{*}{2 materials}	 & variance$_A(cm^2)$	& $2.25 \times 10^{-6}$  &  $9.55 \times 10^{-2}$ & - \\
%\multirow{2}{*}{(3 bins)}  & variance$_B(cm^2)$ & $2.52 \times 10^{-6}$ &  $1.05 \times 10^{-1}$ & -\\
% 		& {MSE $(cm^2)$}	& $2.89 \times 10^{-6}$  &  $1.05 \times 10^{-1}$ & - \\
%\hline		

\multirow{4}{*}{3 materials} & variance$_A(\si{cm^2})$	& $6.11 \times 10^{-6}$  &  $2.67 \times 10^{-1}$ & $3.04 \times 10^{-3}$ \\
  & variance$_B(\si{cm^2})$ &	 $4.82 \times 10^{-6}$  &  $3.33 \times 10^{-1}$ & $2.02 \times 10^{-3}$ \\
& {MSE $(\si{cm^2})$}	& $4.95 \times 10^{-6}$ & $3.33 \times 10^{-1}$ & $2.61 \times 10^{-3}$ \\
%& Bias (cm) & $3.88\times 10^{-4}$ & $1.37 \times 10^{-2}$ & $2.45 \times 10^{-2}$ \\
& Bias (\%) & 3.88 & 0.91 & 11.14\\
		\hline

\multirow{2}{*}{3 materials}	& variance$_B(\si{cm^2})$ &$5.25\times 10^{-6}$ & $3.47 \times 10^{-1}$ &  $2.31 \times 10^{-3}$ \\ 	
\multirow{2}{*}{(Scatter rejected)} & {MSE $(\si{cm^2})$} & $5.27\times 10^{-6}$ & $3.47 \times 10^{-1}$ & $2.31 \times 10^{-3}$ \\ 
%& Bias (cm) & $1.68\times 10^{-4}$ & $1.83 \times 10^{-2}$ & $4.69 \times 10^{-4}$ \\
& Bias (\%) & 1.67 & 1.22 & 0.21\\
	\hline	
\end{tabular}
\label{tab:MSE} }
\end{table*}

\begin{figure}[tbp]
\centerline{ \subfigure[] {\includegraphics[width=0.28\textwidth, angle=-90]{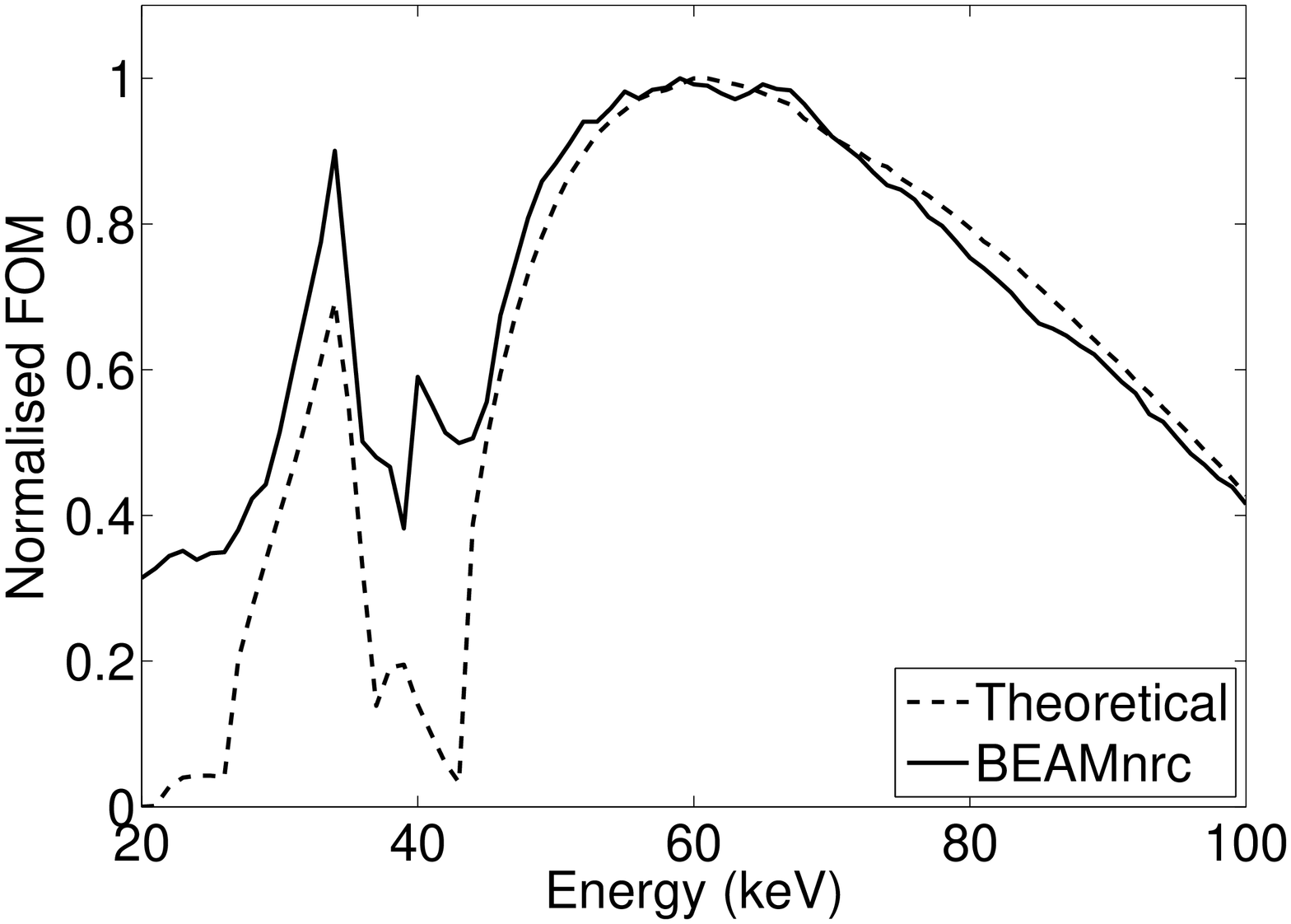}%
\label{fig_FOM220um}}
\hfil
\subfigure[]{ \includegraphics[width=0.28\textwidth, angle=-90]{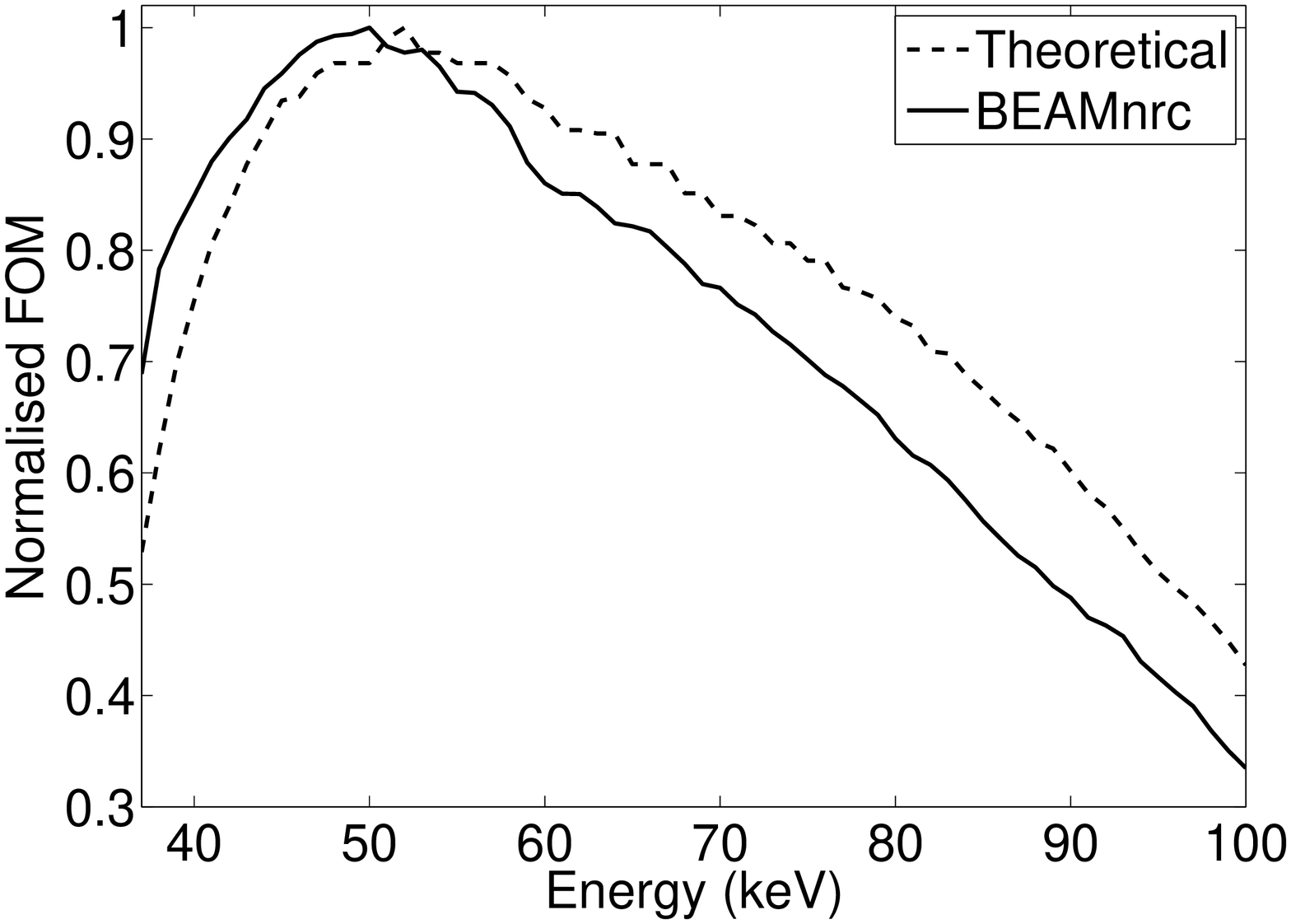} %
\label{fig_FOM_I+Ca+H2O}}}
\caption{Consistencies between the simulated and the theoretical optimal bin border energies. (a) The highest FOM value obtained with the BEAMnrc model differs by \SI{2}{\kilo\electronvolt} from the theoretical optimum of \SI{60}{\kilo\electronvolt} for the decomposition of iodine and water. (b) Likewise, for the decomposition of three materials, optimal $E_{(h,2)}$ was located at \SIlist{51;49}{\kilo\electronvolt} for the theoretical and the BEAMnrc simulation, respectively.}
\label{fig_FOM}
\end{figure}

The FOM curves based on (\ref{eq:fomMSE}) obtained using the BEAMnrc model largely agree with the ones obtained from the optimization algorithm. Figure \ref{fig_FOM} shows the highest FOM value given by the BEAMnrc model is \SI{2}{\kilo\electronvolt} lower than the theoretical optimum at \SI{60}{\kilo\electronvolt} for the decomposition of 0.01\,cm iodine and 1.5\,cm water. Similarly, for three materials, the highest FOM value obtained for the BEAMnrc model was located at \SI{49}{\kilo\electronvolt} compared to \SI{51}{\kilo\electronvolt} for the theoretical optimum. The predicted FOM values for $\pm$\SI{2}{\kilo\electronvolt} around the theoretical optimum was observed to be $>$96\% of the peak value for the BEAMnrc model in both cases. 

\section{Discussion and summary}\label{discussion}
BEAMnrc simulations allow for the optimization of material discrimination to be validated in an idealized environment. No imperfections other than the scattered radiation have been taken into account in the simulations. As shown, optimization of energy bins can provide better confidence in material thickness estimation. While it can be intuitive to place an energy threshold at the K-edge of the imaging material, there may be a more optimal energy, as shown in figure \ref{fig_FOM220um}, due to better counting statistics. For non K-edge imaging, the optimization is particularly crucial to provide an optimal photon binning scheme. Furthermore, some contrast agent with higher atomic number and higher K-edge energy may not be optimal for achieving a balance between contrast and counts. 
%Upon comparing with the simulated variance, the theoretical prediction of image noise presents some limitations. They were identified to be incorporation of image noise sources from both the projection and open beam images, software limitations in the theoretical model, low simulated detected counts relative to clinical settings, and inclusion of scattered radiations. We elaborate on these limitations and describe their potential solutions in the next paragraphs. The simulated $\sigma^2$ incorporated both image noise sources from the projection and the open beam image. The normalization of the spatial variation in photon counts, which was primarily due to the heel effect, introduced a further photon counting noise inherent to the open beam image. 

For the decomposition of two materials in this work, excellent agreement between the predicted and simulated $\sigma^2$ was achieved. A dose calculation procedure, such as \cite{BooneDgN,BooneExposure}, may be implemented on the theoretical model upon the validation to convert the estimated counts into e.g. mean glandular dose required to confidently decompose a calcification feature within breast tissue (see \cite{NikThesis}). For three materials, particularly for calcium, Matlab software limitations precluded a more desirable agreement between predicted and simulated variances. The theoretical prediction of image noise (dotted line) in figure \ref{fig_imageNoiseCa3mat} is limited by the largest possible matrix size and the maximum element in an array allowed in Matlab. This imposed a limit on the step size of the thickness range that could be sampled to form our confidence region, which subsequently hinders the resolution on the change of the size of the confidence region. One potential solution is to run the code on a different platform using \href{http://www.mathworks.com/support/solutions/en/data/1-IHYHFZ/index.html}{a different version of Matlab}. Despite the limitations, the theoretical optima of bin border energies were found accurate to within $\pm$2\,keV, for the discrimination of two and three materials. This has been validated under the consideration of scattered radiation and a realistic scanning geometry. 

\begin{figure}[!t]
\centerline{
\subfigure[] {\includegraphics[width=0.28\textwidth, angle=-90]{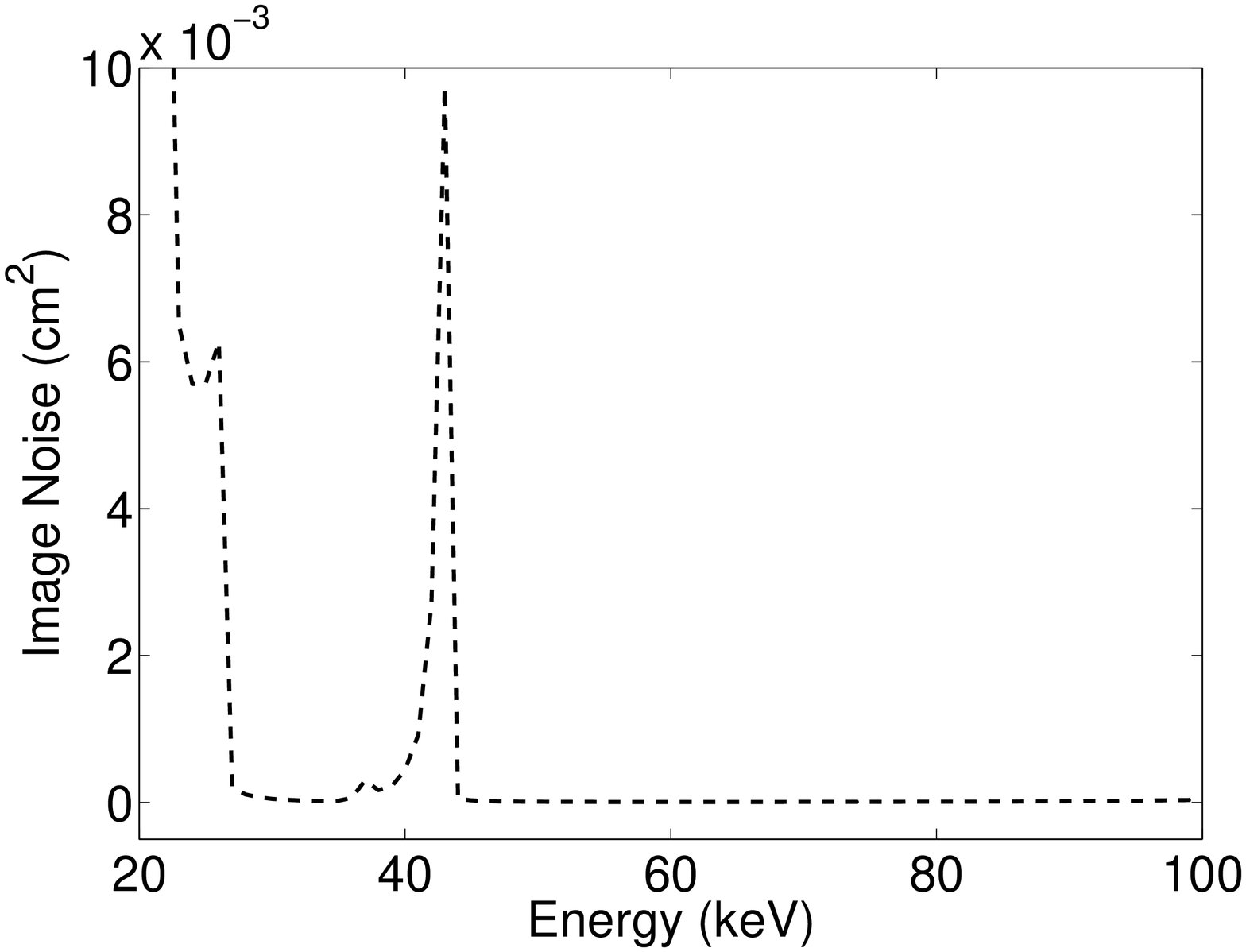}%
\label{fig:IimageTheoryExpand}} 
\subfigure[]{ \includegraphics[width=0.28\textwidth, angle=-90]{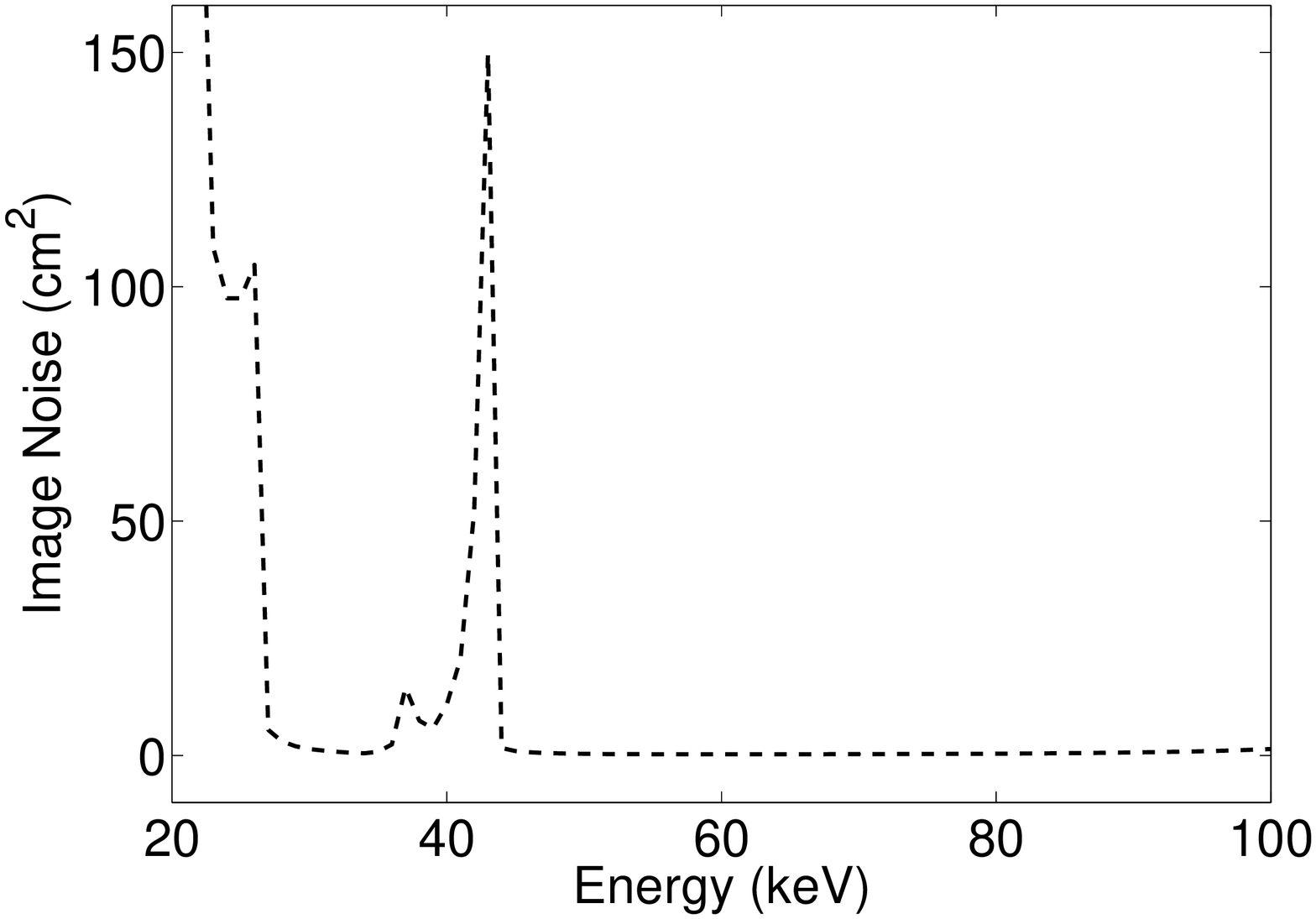}  %
\label{fig:H2OimageTheoryExpand}} }
%\hfil 
%\centerline{ \subfloat[] {\includegraphics[width=0.18\textwidth, angle=-90]{fig_imNoiseITheory}% 
%\label{fig:IimageTheory}}
%\subfloat[] {\includegraphics[width=0.18\textwidth, angle=-90]{fig_imNoiseH2OTheory}%
%\label{fig:H2OimageTheory}} }
\caption{%Image noise $(\sigma^2)$ in the discrimination of iodine/water were reproduced from figure \ref{fig_imageNoise2mat} for (a) iodine and (b) water. 
While figure \protect\ref{fig_imageNoise2mat} focused on the lowest $\sigma^2$ values for (a) iodine and (b) water at the optimal $E_{(h,1)}$ value, the vertical axis was rescaled in this figure to show the entire range of image noise. The confidence region in the theoretical model was allowed to extend indefinitely when the photon counting statistics were poor for a bin border energy. The predicted image noise hence extended beyond the axis in figure \protect\ref{fig_imageNoise2mat}. 
%Note that the $\sigma^2$ values were computed based on the photon flux simulated in the BEAMnrc model.
}
\label{imNoiseTheory2mat}
\end{figure}

Regarding figure \ref{fig_imageNoise2mat}, the confidence region in the theoretical model can expand infinitely when the counting statistics for a bin border energy is poor. The predicted image noise hence extended further than the axis, as shown in figure \ref{fig_imageNoise2mat}. Figure \ref{imNoiseTheory2mat} depicts plots of the entire range of image noise.

%Although the BEAMnrc model was setup in this work to enable separate simulations of x-ray tube and transmission through the imaging object, the generation of x-ray photons was largely inefficient, particularly for tube voltages comparatively lower than 120\,kVp. Only the tungsten spectrum of 120\,kVp was simulated in this work, as a result. Furthermore, the calculation time of the in-house Matlab code to translate phsp information into projection image increased proportionally to the number of incident particles. The pixel size on the detector plane was set to be four times that of the Medipix detector to partially compensate for these limitations. For the given imaging geometry, object size and the number of primary history, it was found that the detector pitch of \SI{220}{\um} achieved an adequate balance between simulated image resolution and spectral SNR for acceptable accuracy in thickness estimation. As such, spectra of smaller pixel pitches can be simulated too. 

It should be noted that, for computational efficiency, the simulations were performed below the typical clinical settings of standard x-ray photon flux rates. Simulated detected counts were less than 900 per pixel for all cases. It is expected that increasing the number of detected counts can facilitate noise reduction in the simulated spectrum %(solid curve) in figure \ref{fig_singlepxVSmean} 
and thereby provide improved agreements between variance$_{A}$ and variance$_{B}$. 
Furthermore, the scatter contribution between \SIlist{10;60}{\kilo\electronvolt} for the three material decomposition was 25\% of the total photon counts, which contributed to the 11\% bias in calcium thickness estimation in table \ref{tab:MSE}. The 10\% scattered radiation between \SIlist{10;60}{\kilo\electronvolt} for the decomposition of two materials does not result in a considerable bias in thickness estimation (variance $\approx$ MSE) and was thus considered negligible. 
While the rejection of scattered radiation lowered the bias in the decomposition, the reduction in simulated detected photon counts resulted in a marginally higher image noise in the decomposition of three materials. 
The quantification and rejection of scattered radiation was enabled by the particle interaction tracking ability in BEAMnrc \cite{BEAMnrcManual}. 
Note that practical implementation of scatter rejection, such as a multi-slit collimators have been implemented by other groups \cite{DingBreast,ShikhalievFeasibility}. 
A future application is therefore scatter correction utilizing the particle tracking function in BEAMnrc, which may help reducing the impact of scattered radiation on material decomposition using spectral x-ray imaging \cite{WiegertScattered}. 

In conclusion, a thorough analysis on the simulated noise was performed and compared with the theoretical prediction to provide a validation of the optimization algorithm in \cite{NikOptimal} without the technical complications of a PCD. Excellent agreement was found between the predicted and simulated image noise for the decomposition of two materials. The prediction of image noise for the decomposition of three materials was impeded by the largest possible matrix size allowed in Matlab. However, the theoretical model was shown to be accurate to within $\pm$2\,keV for the discrimination of two and three materials. Scattered radiation was shown to only minimally affect the optimal bin borders. The validated model can also be implemented to estimate the counts per pixel needed for achieving a specific imaging aim in the decomposition of two materials, such as to conÞdently decompose a calciÞcation feature within breast tissue.

\acknowledgments
We gratefully acknowledge Tony Teke (BC Cancer Agency, Vancouver, Canada) for providing the initial code to read the BEAMnrc phsp files, and Dr. Vladimir Mencl and Dr. Fran{\c c}ois Bissey for their help on the BlueFern$\textsuperscript{\textregistered}$ setup. 
SJN would like to thank \href{http://www.marsbioimaging.com}{Mars Bioimaging Ltd (MBI)} for his PhD scholarship. RST would like to acknowledge a Study Abroad Grant from Nordeafonden in Copenhagen and the University of Canterbury summer scholarship scheme for their financial support throughout this work. 

\bibliographystyle{jhep}
\bibliography{MC_validation_arxiv}

\end{document}